\documentclass[preprint,amsmath,amssymb,aps,a4]{revtex4-2}
\usepackage[colorlinks=true, citecolor=blue, linkcolor=blue, urlcolor=blue]{hyperref}
\usepackage{xcolor}
\usepackage{graphicx}
\usepackage{amssymb}
\usepackage{multirow}
\usepackage{array}
\usepackage{float}
\usepackage{cleveref}
\usepackage{comment}
\usepackage{xcolor}
\usepackage{amsfonts}
\usepackage{rotating}
\usepackage{xcolor}
\usepackage{multirow}
\usepackage{array}
\usepackage{makecell}
\usepackage{amsmath,amssymb}
\usepackage{subfig}
\usepackage{booktabs}
\usepackage{array}

\hypersetup{
    colorlinks=true,
    citecolor=blue
}

\usepackage[justification=raggedright,singlelinecheck=false]{caption}

\begin{document}
	
\title{Pion structure in magnetized nuclear medium at finite temperature}

\author{Fathimath Shifa N K}
\email{fathimathshifa100101@gmail.com}
\affiliation{Department of Physics, Dr. B. R. Ambedkar National
		Institute of Technology, Jalandhar, 144008, India}
	
\author{Navpreet Kaur}
\email{knavpreet.hep@gmail.com}
\affiliation{Department of Physics, Dr. B. R. Ambedkar National
		Institute of Technology, Jalandhar, 144008, India}

\author{Abi Jebarson A}
\email{abijebarson@gmail.com}
\affiliation{Department of Physics, Dr. B. R. Ambedkar National
		Institute of Technology, Jalandhar, 144008, India}

\author{Arvind Kumar}
\email{kumara@nitj.ac.in}
\affiliation{Department of Physics, Dr. B. R. Ambedkar National
		Institute of Technology, Jalandhar, 144008, India}

\author{Suneel Dutt}
\email{dutts@nitj.ac.in}
\affiliation{Department of Physics, Dr. B. R. Ambedkar National
		Institute of Technology, Jalandhar, 144008, India}
	
\author{Harleen Dahiya}
\email{dahiyah@nitj.ac.in}
\affiliation{Department of Physics, Dr. B. R. Ambedkar National
		Institute of Technology, Jalandhar, 144008, India}
	
\date{\today}%
\newpage

\begin{abstract}
We investigate the structure of the pion in a magnetized nuclear medium at finite temperature by analyzing the in-medium modifications of its parton distribution functions (PDFs) and electromagnetic form factors (EMFFs). We use a unified framework combining the chiral SU(3) quark mean field (CQMF) model and the light-cone quark model (LCQM) where the in-medium quark masses are evaluated using the CQMF model and subsequently employed as inputs in the light-cone formalism. The study systematically explores the effects of magnetic field, baryon density, temperature, and isospin asymmetry on the pion structure. In addition, Dirac sea contributions are incorporated to account for vacuum polarization effects in the medium. Our results of PDFs and EMFFs provide valuable insights into the modification of pion properties under extreme conditions, highlighting the significant role of the surrounding medium in altering its internal structure. 
\end{abstract}
%
%====================================================
%
\maketitle
\newpage
\section{Introduction}
The modification of the vacuum properties of hadrons due to their surrounding medium is one of the most fascinating topics in contemporary high-energy physics. These medium-induced changes can be reflected in measurable observables such as form factors and cross sections \cite{EuropeanMuon:1983wih,Suzuki:2002ae}. Consequently, these observables provide valuable information about the internal structure and dynamics of hadrons in the nuclear medium \cite{Hayano:2008vn, Hosaka:2016ypm, Leupold:2009kz,Post:2003hu, Metag:2017yuh,Tolos:2020aln}. Experimentally, it has been established that the structure functions of nucleons are modified inside nuclei, a phenomenon known as the European Muon Collaboration (EMC) effect \cite{EuropeanMuon:1983wih}. Similar medium-induced modifications are also expected in other hadrons, along with changes in their effective masses, width broadening, and an increase in their charge radii \cite{Montesinos:2024xpp,KEK-PS-E325:2005wbm, JeffersonLabE93-049:2002asn, Suzuki:2002ae}, when they are embedded in a nuclear medium. These phenomena are closely associated with the partial restoration of chiral symmetry, which has been experimentally \cite{Suzuki:2002ae, Sgaramella:2024qdh, Friedman:2004jh, CHAOS:1996nql, CHAOS:2004rhl} as well as theoretically \cite{Itahashi:2024apl, Gifari:2024ssz,Hutauruk:2018qku, Jido:2008bk} supported. Understanding the in-medium modification of hadron properties is also crucial for interpreting the dilepton spectra arising from vector meson decays \cite{Fuchs:2004uu}. These theoretical investigations are of considerable importance for ongoing and future experiments, including the PANDA and CBM collaborations at FAIR \cite{PANDA:2009yku, Dbeyssi:2022zwz, Wilczek:2010ae, CBM:2016kpk, Prencipe:2015cgg}, the PHENIX experiment at RHIC \cite{PHENIX:2005nhb}, the Belle and Belle II experiments at KEK \cite{Prencipe:2018ugj}, the LHCb experiment at CERN \cite{LHCb:2011zzp}, J-PARC in Japan \cite{Kumano:2022cje, Aoki:2021cqa}, NICA at JINR \cite{Sissakian:2009zza} and the Electron-Ion Collider (EIC) in the United States \cite{AbdulKhalek:2021gbh}.
\par
%As the lightest meson and the Goldstone boson associated with the spontaneous breaking of chiral symmetry, the pion plays a fundamental role in understanding non-perturbative QCD. 
The pion, being the lightest meson and the Goldstone boson associated with the spontaneous breaking of chiral symmetry plays a fundamental role in understanding non-perturbative quantum chromodynamics (QCD). The vacuum structure and properties of the pion have been extensively studied from both theoretical and experimental perspectives. Although the absence of free pion targets poses a significant challenge for the direct experimental investigation of its internal structure, valuable information has been obtained through processes involving pion exchange and pion production. The primary source of experimental information on the pion PDFs has been pion-induced Drell-Yan measurements \cite{NA3:1983ejh,E615:1989bda,H1:2010hym}. However, the future tagged pion exchange (Sullivan) process at the EIC \cite{Aguilar:2019teb}, pion-induced hard exclusive reactions ($\pi^{-}p \rightarrow \gamma^{*}n$) at the COMPASS and AMBER experiments \cite{Adams:2018pwt}, and exclusive pion electroproduction ($ep \rightarrow e^{\prime}\pi^{+}n$) at the upgraded 12 GeV Jefferson Lab \cite{Arrington:2021alx} are expected to provide important constraints on the pion structure. On the theoretical side, the internal structure of the pion has been extensively investigated through observables such as the parton distribution functions (PDFs) and electromagnetic form factors (EMFFs) \cite{Alexandrou:2026nsl, Francis:2025rya, Bednar:2018mtf, Ding:2019qlr, Chen:2016sno, Hecht:2000xa}. The PDFs describe the distribution of the longitudinal momentum fraction $x$ carried by the quark inside the pion \cite{Collins:1981uw,Sharma:2016cnf,Martin:2009iq}, whereas the EMFFs encode the spatial distributions of charge and magnetization \cite{Perdrisat:2006hj,Diehl:2013xca,Khodjamirian:2006st}. Since both the PDFs and EMFFs are directly related to the underlying quark dynamics, they provide valuable insights into the internal structure of the pion and serve as sensitive probes of medium-induced modifications.
\par In addition to studying pion properties in vacuum, many studies have been carried out to investigate the in-medium modification of various mesonic properties, such as form factors, distribution amplitudes (DAs), weak decay constants, and others. These studies have been performed using different theoretical approaches, including the quark-meson coupling (QMC) model~\cite{Arifi:2024tix,Kim:2026kkf,Zeminiani:2020aho}, the Dyson-Schwinger equation (DSE) approach \cite{Albino:2022gzs,Chang:2011vu,Fischer:2018sdj}, QCD Sum Rules (QSR) \cite{Park:2016xrw,Bozkir:2022lyk}, the linear sigma model (L$\sigma$M) \cite{Suenaga:2019urn}, the Bethe-Salpeter equation-Nambu-Jona-Lasinio (BSE-NJL) model \cite{Hutauruk:2018qku,Hutauruk:2021kej,Hutauruk:2019ipp,Hutauruk:2019was}, and the instanton liquid model (ILM) \cite{Nam:2008xx,Shuryak:1997vd}. Hybrid approaches, such as the light-front constituent quark model (LFCQM) with quark-meson coupling (LFCQM-QMC) \cite{deMelo:2016uwj,deMelo:2014gea,deMelo:2018hfw} and the light-front quark model implemented with quark-meson coupling model (LFQM-QMC) \cite{Arifi:2024tix,Arifi:2023jfe}, have also been employed in these studies. The in-medium structure and properties of hadrons have been extensively investigated for various nuclear matter parameters, including finite baryon density, temperature, and isospin asymmetry \cite{Hutauruk:2021kej,Arifi:2023jfe,Arifi:2024tix,Hutauruk:2018qku,Hutauruk:2019was,Hutauruk:2019ipp,Yabusaki:2023zin,Er:2022cxx,Puhan:2024xdq,Kaur:2024wze,Singh:2024lra}. In particular, the in-medium properties of the pion in symmetric nuclear matter have been studied by combining the LF framework with the QMC model for different baryon densities~\cite{deMelo:2014gea}. Subsequently, the effects of finite temperature and isospin asymmetry on the pion structure were investigated within the light-cone quark model (LCQM), employing the in-medium quark masses obtained from the chiral SU(3) quark mean field (CQMF) model ~\cite{Puhan:2025ibn,Puhan:2024xdq}.

The combined CQMF-LCQM framework provides a self-consistent approach for investigating the in-medium structure of hadrons by connecting the medium-modified constituent quark masses obtained from the CQMF model with the relativistic light-front description of hadrons in the LCQM. The CQMF model treats hadrons at the constituent quark level and provides a unified description of strongly interacting matter under different conditions of baryon density, temperature, isospin asymmetry, and magnetic fields while incorporating the essential low-energy features of QCD, such as chiral symmetry and its spontaneous breaking \cite{Wang:2001hw,Wang:2001jw,Wang:2004wja,Mukherjee:2018ebw}. On the other hand, the LCQM provides a relativistic and gauge-invariant framework for describing the internal structure of hadrons in terms of their valence quark degrees of freedom. It has been successfully employed to describe a wide range of hadronic observables, including EMFFs, charge radii, decay constants, PDFs, DAs, and transverse momentum-dependent parton distributions (TMDs), with good agreement with available experimental and lattice QCD results \cite{Kaur:2019jow, Puhan:2024xdq, P:2026crg}. However, despite these studies, the in-medium modification of the pion structure in hot magnetized nuclear matter has not yet been systematically investigated within the combined CQMF and LCQM framework.
%
%
%
%
%%
%
%
%
%
%
%
%The existing studies have considerably improved our understanding of the medium modification of pion structure.
%
\par In particular, strong magnetic fields are generated in non-central relativistic heavy-ion collisions, reaching strengths of approximately $eB \sim 2\,m_{\pi}^{2}$ at RHIC and $eB \sim 15\,m_{\pi}^{2}$ at the LHC \cite{Kharzeev:2007jp,Fukushima:2008xe}. 
Further, the inclusion of magnetic fields therefore provides a more comprehensive description of the nuclear environment as they are expected to exist in various astrophysical objects, such as neutron stars and magnetars, and are also believed to have been present during the evolution of the early universe \cite{Turolla:2015mwa,Kumari:2022jvq,Kundu:2022nva,Vachaspati:1991nm,Durrer:2013pga}. It has also been found that the presence of a magnetic field gives rise to several novel phenomena in strongly interacting matter, including magnetic catalysis (MC), inverse magnetic catalysis (IMC), and the chiral magnetic effect (CME) \cite{Gusynin:1995nb,Bali:2012zg,Huang:2022qdn}. The MC and IMC under different medium conditions have been briefly studied in Refs.~\cite{Haber:2014ula,Li:2016gfn,Fang:2016cnt}. Furthermore, the inclusion of magnetized Dirac sea (DS) contributions has been shown along with its role in the recognition of MC and IMC, thereby influencing the in-medium properties of hadrons. In addition, the anomalous magnetic moments (AMMs) of nucleons have also been incorporated in several studies as they significantly affect the behavior of magnetized nuclear matter and play a crucial role in realization of IMC and MC at finite baryon density \cite{Mishra:2023uhx,Mukherjee:2018ebw}.

Given the strong magnetic fields expected in relativistic heavy-ion collisions, in the present work, we investigate the in-medium structure of the pion in hot magnetized nuclear matter for both symmetric and asymmetric nuclear matter within a unified framework combining the CQMF model and LCQM. The in-medium quark masses obtained from the CQMF model are employed as inputs to the LCQM to evaluate the in-medium PDFs and EMFFs of the pion. In the present formalism, the magnetic field is incorporated only at the nucleon level. Consequently, the pion experiences the influence of the magnetic field indirectly through the medium-modified quark masses rather than through a direct coupling of the magnetic field to its constituent quarks. Furthermore, the effects of magnetized DS contributions are systematically investigated by comparing the results obtained with and without DS, thereby highlighting the role of vacuum polarization in the in-medium modification of the pion structure.
\par This paper is organized as follows. Section~\ref{sec:Formalism} begins with a brief description of the CQMF model and its implementation for evaluating the in-medium quark masses, which serve as inputs to the LCQM. It also presents the LCQM formalism, including the expressions for the PDFs and EMFFs. The results of the present work are discussed in Sec.~\ref{sec:Results and Discussion}, and the main conclusions are summarized in Sec.~\ref{sec:Summary and Conclusions}.
\section{Formalism}
\label{sec:Formalism}
A unified framework combining CQMF and LCQM is employed to investigate the in-medium pion properties. This section focuses on the evaluation of effective quark masses in the medium and the subsequent computation of medium-induced modifications to the pion structure through a detailed description of the underlying formalism.
\subsection{Chiral SU(3) Quark Mean Field (CQMF) Model}
%
%
%
%In CQMF model, quark-meson and meson-meson interactions are considered using a nonlinear realization of chiral SU(3) symmetry \cite{Weinberg:1968de,Coleman:1969sm,Bardeen:1969ra} together with broken scale invariance \cite{Papazoglou:1998vr,Mishra:2003se,Mishra:2003tr}, extended to finite temperature and density. 
In CQMF model, quark-meson and meson-meson interactions are incorporated by considering the low energy properties of QCD such as spontaneous and explicit breaking of chiral symmetry as well as the broken scale invariance  \cite{Papazoglou:1998vr,Wang:2001hw,Wang:2001jw,Wang:2004wja}.
Within hadrons, quarks are confined and their dynamics are governed by interactions mediated through scalar and vector meson fields. The scalar fields considered are $\mathrm{\sigma}$, $\mathrm{\zeta}$, and $\mathrm{\delta}$, along with the scalar dilaton field $\mathrm{\chi}$, whereas the vector fields include $\mathrm{\omega}$ and $\mathrm{\rho}$. Among these, the $\mathrm{\delta}$ and $\mathrm{\rho}$ fields play a crucial role in describing the isospin asymmetry of the medium. The dynamics and interactions of these fields are systematically described by the Lagrangian density of the model, which is given by \cite{Kumari:2020mci}
\begin{align}
\mathcal{L}_{\text{CQMF}} = \mathcal{L}_{q 0} + \mathcal{L}_{qM} + \mathcal{L}_{VV}+ \mathcal{L}_{\Sigma\Sigma }  + \mathcal{L}_{\chi SB}  + \mathcal{L}_{CP}.
%+ \mathcal{L}_{\Delta m}.
\end{align}
%
%%
%%%%%%%%%%%
Here, each term represents a distinct physical contribution. The term $\mathcal{L}_{q0}$ corresponds to the kinetic energy of free quarks, while $\mathcal{L}_{q M}$ describes the interaction of quarks with meson fields. The strong interaction among quarks is modeled through the exchange of scalar and vector meson fields. The term $\mathcal{L}_{VV}$ governs the dynamics of vector mesons such as $\omega$ and $\rho$ \cite{Wang:2001hw}, whereas $\mathcal{L}_{\Sigma\Sigma}$ accounts for scalar meson interactions, including effects associated with spontaneous breaking of chiral symmetry and scale invariance. The explicit breaking of chiral symmetry, which generates masses for pseudoscalar mesons, is incorporated through $\mathcal{L}_{\chi SB}$. The term $\mathcal{L}_{CP}$ represents an effective confining potential that ensures quark confinement within hadrons. %Additionally, $\mathcal{L}_{\Delta m}$ accounts for mass contributions and incorporates a finite strange quark mass through explicit symmetry breaking \cite{Wang:2001jw,Singh:2016hiw}. 
The detailed expressions of these terms can be found in Refs.~\cite{Kumari:2020mci,Puhan:2024xdq,Wang:2001hw,Wang:2001jw,Wang:2004wja}.
An additional term, $\mathcal{L}_{\text{mag}}$, is introduced to account for the effects of the external magnetic field at the nucleon level, leading to the total effective Lagrangian \cite{Mishra:2023uhx, Mukherjee:2018ebw} as
\begin{align}
\mathcal{L}_{\text{tot}} = \mathcal{L}_{\text{CQMF}} + \mathcal{L}_{\text{mag},}
\end{align}
in which, 
\begin{equation}
\mathcal{L}_{\mathrm{mag}} 
= - \bar{\psi}_i q_i \gamma^\mu A_\mu \psi_i 
- \frac{1}{2} \kappa_i \bar{\psi}_i \sigma^{\mu\nu} F_{\mu\nu} \psi_i 
- \frac{1}{4} F^{\mu\nu} F_{\mu\nu}.
\label{eq:Lmag}
\end{equation}
%
%
%
%
%
%where $\mu_i$ represents the intrinsic magnetic moment of the $i^{\text{th}}$ nucleon,
Here, $q_i$ represents the electric charge of the $i^{\mathrm{th}}$ nucleon described by the field $\psi_i$. The second term corresponds to a tensor-type interaction between nucleons and the electromagnetic field, formulated through the field strength tensor $F^{\mu\nu} = \partial^\mu A^\nu - \partial^\nu A^\mu$ and $\sigma^{\mu\nu} = \frac{i}{2}[\gamma^\mu, \gamma^\nu]$. The parameter $\kappa_i$ denotes the anomalous magnetic moment (AMM) of the $i^{\mathrm{th}}$ nucleon in terms of $\mu_N$, $q_i$ is the charge and $\mathcal{M}_p$ is the proton mass. In the presence of a constant magnetic field, this interaction becomes significant. For charged particles like proton, the Lorentz force leads to a quantization of the transverse motion, resulting in discrete Landau levels. The emergence of Landau levels modifies the energy spectrum of nucleons in the medium and consequently affects the thermodynamic properties of magnetized nuclear matter. In particular, the thermodynamic potential receives separate contributions corresponding to motion parallel and perpendicular to the magnetic field, with distinct behavior for proton and neutron. The modified thermodynamic potential can be expressed as \cite{Mishra:2023uhx}
\begin{align}
\Omega = \Omega_{med} + \Omega_{DS} - \mathcal{L}_{VV}- \mathcal{L}_{\Sigma\Sigma}  - \mathcal{L}_{\chi SB}.
\label{eq:total_potential}
\end{align}
The quantities $\Omega_{\text{med}}$ and $\Omega_{DS}$ represent the contributions from the Fermi sea (FS) and DS to the thermodynamic potential, respectively. For baryons with spin $1/2$, the thermodynamic potential for charged baryons, such as protons, takes the form \cite{Haber:2014ula,Aguirre:2016vqa,Aguirre:2019ivr}
\begin{align}
\Omega_{DS}^{p} &= - \frac{|q_p| B}{2\pi} \Bigg[ \Bigg( \sum_{\nu=0}^{\nu_{\text{max}}} 
\int_{0}^{\infty} \frac{dk_\parallel}{2\pi} \, \tilde{E}_{k,\nu,s}^p\Bigg)_{s=+1}+\Bigg(\sum_{\nu=1}^{\nu_{\text{max}}} 
\int_{0}^{\infty} \frac{dk_\parallel}{2\pi} \, \tilde{E}_{k,\nu,s}^p \Bigg)_{s=-1}\Bigg], 
\end{align}
and
\begin{align}
\Omega^{p}_{\mathrm{med}} 
&= -T \, \frac{|q_p| B}{2\pi} 
\Bigg[ \Bigg(\sum_{\nu=0}^{\nu_{\text{max}}}  
\int_{0}^{\infty} \frac{dk_{\parallel}}{2\pi}  \times \left\{
\ln\left(1 + e^{-\beta(\tilde{E}_{k,\nu,s}^p - \mu_p^*)}\right)
+ \ln\left(1 + e^{-\beta(\tilde{E}_{k,\nu,s}^p + \mu_p^*)}\right)
\right\} \Bigg)_{s=+1}\notag \\
& +\Bigg(\sum_{\nu=1}^{\nu_{\text{max}}}  
\int_{0}^{\infty} \frac{dk_{\parallel}}{2\pi}  \times \left\{
\ln\left(1 + e^{-\beta(\tilde{E}_{k,\nu,s}^p - \mu_p^*)}\right)
+ \ln\left(1 + e^{-\beta(\tilde{E}_{k,\nu,s}^p + \mu_p^*)}\right)
\right\} \Bigg)_{s=-1} \Bigg].
\end{align}
Here, $\beta = 1/T$, $\mu_p^*$ is the effective chemical potential of the proton, and $|q_p| = e$ is the proton charge. Furthermore, $B$ denotes the external magnetic field,  $s $ denotes the orientation of the nucleon spin relative to the magnetic field, and $\nu$ and $\nu_{\mathrm{max}}$ represent the Landau level and the maximum occupied Landau level, respectively. For neutral baryons, such as neutrons, the DS and medium contributions to the thermodynamic potential are expressed as \cite{Haber:2014ula,Aguirre:2016vqa,Aguirre:2019ivr}
\begin{align}
\Omega^{n}_{DS} = - \sum_{s=\pm 1} \int \frac{d^3 k}{(2\pi)^3} \, \tilde{E}_{k,s}^n,
\end{align}
and
\begin{align}
\Omega^{n}_{\text{med}}= & -T \sum_{s=\pm 1} \int \frac{d^3k}{(2\pi)^3}  \left[
\ln\left(1 + e^{-\beta(\tilde{E}^n_{k,s} - \mu_n^*)}\right)
+ \ln\left(1 + e^{-\beta(\tilde{E}^n_{k,s} + \mu_n^*)}\right)
\right],
\end{align}
respectively, where $\mu^*_n$ represent the chemical potential of neutron. The single-particle energies of nucleons are modified in the presence of an external magnetic field. The corresponding energy for protons is given by \cite{Mishra:2023uhx, Mukherjee:2018ebw}
\begin{align}
\tilde{E}_{k,\nu,s}^p = \sqrt{
k_\parallel^2 + (\bar{m}^p_{\nu,s})^2
},
\label{eq:charged_energy}
\end{align}
where $\bar{m}^{p}_{\nu,s}=\sqrt{2\nu |q_i| B + \mathcal{M}_p^{*2}} - s \kappa_p B$. The effective energy of a neutron in an external magnetic field is given by
\begin{align}
\tilde{E}_{k,s}^n = \sqrt{(k_{\parallel})^2 + (\bar{m}^n_{k,s})^2},
\end{align}
with $k_{\perp}$ and $k_{\parallel}$ denote the transverse and longitudinal momentum components with respect to the direction of the magnetic field and $\bar{m}^n_{k,s}=\sqrt{\mathcal{M}_n^{*2} + (k_\perp)^2} - s \kappa_n B$. The effective mass of the $i^{\mathrm{th}}$ nucleon in the medium can be expressed as \cite{Barik:1985rm, Barik:2013lna}
\begin{align}
\mathcal{M}_i^* = \sqrt{E_i^{*2} - \langle p_{i_{\text{cm}}}^{*2} \rangle}. 
\end{align}
Here, $E_i^*$ represents the in-medium effective energy of the corresponding nucleon and can be expanded as 
\begin{align}
E_i^* = \sum_q n_{qi} e_q^* + E_{i,\text{spin}},
\end{align}
with, $n_{qi}$ as the number of quarks of type $q$ ($q=u,d$) in the nucleon, and $e_q^*$ represents the effective energy of the corresponding quark. The term $E_{i,\text{spin}}$ corresponds to the spin-spin interaction, which is adjusted to reproduce the nucleon masses in free space. The quantity $\langle p_{i_{\text{cm}}}^{*2} \rangle$ represents the expectation value of the squared center-of-mass momentum, accounting for the spurious motion of the center of mass in describing the nucleon as a bound system of quarks, and can be written as
\begin{align}
     \langle p_{i_{\text{cm}}}^{*2} \rangle= \sum_q  \langle p_{i,\text{cm}}^{*2} \rangle_q=\sum_q \frac{11e_q^*+m_q^*}{6(3e_q^*+m_q^*)}(e_q^{*2}-m_q^{*2}),
\end{align}
where quantity $m_{q}^*$ represents the effective quark mass, which takes the form
\begin{align}
m_q^* = -g_\sigma^q \sigma - g_\zeta^q \zeta - g_\delta^q I^{q}_{3} \delta ,
%+ m_{q0},
\label{eq:mod_mass}
\end{align}
where $g_\mathrm{\sigma}^q$, $g_\mathrm{\zeta}^q$ and $g_\mathrm{\delta}^q$ are the coupling constants associated with the respective scalar meson fields $\mathrm{\sigma}$, $\mathrm{\zeta}$ and $\mathrm{\delta}$. 
%The parameter $m_{q0}$, representing the vacuum quark mass, is assigned a finite value for the strange quark, while it is set to zero for the up and down quarks.
$I^{q}_{3}$ is the third isospin component which is $+\frac{1}{2}$ for $u$ quark and $-\frac{1}{2}$ for $d$ quark. The quark masses in vacuum can be computed from the expectation values of the fields in vacuum, which is given by
%
%
% \begin{equations}
% \begin{align}
% m_u &= m_d=-\frac{g_s}{\sqrt{2}}\mathrm{\sigma_0}=-g_\mathrm{\sigma}^q\ \mathrm{\sigma_0}.,  \\
% m_s &= -g_\zeta^s \zeta_0 + m_{s0}.
% \end{align}
% \end{equations}%
\begin{eqnarray}
    m_u &= m_d=-\frac{g_s}{\sqrt{2}}\mathrm{\sigma_0}=-g_\mathrm{\sigma}^q\ \mathrm{\sigma_0}.
\end{eqnarray}
At specified values of temperature $T$, baryon density $\rho_B$, isospin asymmetry $\eta$, and magnetic field $eB$, the scalar fields ($\sigma$, $\zeta$, and $\delta$), along with the dilaton field ($\chi$) and the vector fields ($\omega$ and $\rho$)  %$\phi$)
, are derived by minimizing the thermodynamic potential $\Omega$ \cite{Kumari:2020mci, Kumar:2019tiw, Kumar:2018ujk}.%
\begin{align}
\frac{\partial \Omega}{\partial \mathrm{\sigma}} =
\frac{\partial \Omega}{\partial \mathrm{\zeta}} =
\frac{\partial \Omega}{\partial \mathrm{\delta}} =
\frac{\partial \Omega}{\partial \chi} =
\frac{\partial \Omega}{\partial \mathrm{\omega}} =
\frac{\partial \Omega}{\partial \mathrm{\rho}} =0.
\end{align}%
The following coupled nonlinear equations are obtained as a result of the minimization,
\begin{align}
\frac{\partial \Omega}{\partial \mathrm{\sigma}} &= 
k_0 \chi^2 \mathrm{\sigma} - 4k_1 (\mathrm{\sigma}^2 + \mathrm{\zeta}^2 + \mathrm{\delta}^2)\mathrm{\sigma} 
- 2k_2 (\mathrm{\sigma}^3 + 3\mathrm{\sigma}\mathrm{\delta}^2) - 2k_3 \chi \mathrm{\sigma} \mathrm{\zeta} - \frac{d}{3} \chi^4 \left( \frac{2\mathrm{\sigma}}{\mathrm{\sigma}^2 - \mathrm{\delta}^2} \right) \nonumber \\
&+ \left( \frac{\chi}{\chi_0} \right)^2 m_\pi^2 f_\pi \quad - \left( \frac{\chi}{\chi_0} \right)^2 m_\omega \mathrm{\omega^2} \frac{\partial m_\omega}{\partial \mathrm{\sigma}} - \left( \frac{\chi}{\chi_0} \right)^2 m_\rho \mathrm{\rho^2} \frac{\partial m_\rho}{\partial \mathrm{\sigma}}
- \sum_{i} g_\sigma^i \rho^\text{S}_i = 0, 
\end{align}
\begin{align}
\frac{\partial \Omega}{\partial \mathrm{\zeta}} &=
k_0 \chi^2 \mathrm{\zeta} - 4k_1 (\mathrm{\sigma}^2 + \mathrm{\zeta}^2 + \mathrm{\delta}^2)\mathrm{\zeta} 
- 4k_2 \mathrm{\zeta}^3 - k_3 \chi (\mathrm{\sigma}^2 - \mathrm{\delta}^2)  - \frac{d}{3} \frac{\chi^4}{\mathrm{\zeta}} + \left( \frac{\chi}{\chi_0}  \right)^2 \nonumber \\
& \times 
\left( \sqrt{2} m_K^2 f_K - \frac{1}{\sqrt{2}} m_\pi^2 f_\pi \right) -\sum_{i} g_\zeta^i \rho^\text{S}_i = 0,
\end{align}
\begin{align}
\frac{\partial \Omega}{\partial \mathrm{\delta}} &= 
k_0 \chi^2 \mathrm{\delta} - 4k_1 (\mathrm{\sigma}^2 + \mathrm{\zeta}^2 + \mathrm{\delta}^2)\mathrm{\delta} 
- 2k_2 (\mathrm{\delta}^3 + 3\mathrm{\sigma}^2 \mathrm{\delta}) + 2k_3 \chi \mathrm{\delta} \mathrm{\zeta} + \frac{2}{3} d \chi^4 \left( \frac{\mathrm{\delta}}{\mathrm{\sigma}^2 - \mathrm{\delta}^2} \right) \nonumber \\
&- \sum_{i} g_\delta^i \tau_3^i\rho^\text{S}_i = 0,
\end{align}
\begin{align}
\frac{\partial \Omega}{\partial \chi} &= 
k_0 \chi (\mathrm{\sigma}^2 + \mathrm{\zeta}^2 + \mathrm{\delta}^2)
- k_3 (\mathrm{\sigma}^2 - \mathrm{\delta}^2)\mathrm{\zeta}
+ \chi^3 \left[ 1 + \ln \left( \frac{\chi^4}{\chi_0^4} \right) \right]
+ (4k_4 - d)\chi^3 \nonumber \\
& - \frac{4}{3} d \chi^3 
\ln \left( \frac{(\mathrm{\sigma}^2 - \mathrm{\delta}^2)\mathrm{\zeta}}{\sigma_0^2 \zeta_0} \left( \frac{\chi}{\chi_0} \right)^3 \right) + \frac{2\chi}{\chi_0^2} 
\left[ m_\pi^2 f_\pi \mathrm{\sigma}
+ \left( \sqrt{2} m_K^2 f_K - \frac{1}{\sqrt{2}} m_\pi^2 f_\pi \right)\mathrm{\zeta} \right] \nonumber \\
& - \frac{\chi}{\chi_0^2} \left( m_\omega^2 \mathrm{\omega}^2 + m_\rho^2 \mathrm{\rho}^2 \right) = 0,
\end{align}
\begin{align}
& \quad \quad \quad \quad \quad \quad \frac{\partial \Omega}{\partial \mathrm{\omega}} = 
\Bigg(\frac{\chi}{\chi_0}\Bigg)^2 m_\omega^2 \mathrm{\omega}
+ 4g_4 \mathrm{\omega}^3
+ 12g_4 \mathrm{\omega}\mathrm{\rho}^2
- \sum_{i} g_\omega^i \rho_i^v = 0,
\end{align}
\begin{align}
& \quad \quad \quad \quad \quad \quad \frac{\partial \Omega}{\partial \mathrm{\rho}} = 
\Bigg(\frac{\chi}{\chi_0}\Bigg)^2 m_\rho^2 \mathrm{\rho}
+ 4g_4 \mathrm{\rho}^3
+ 12g_4 \mathrm{\omega}^2 \mathrm{\rho}
- \sum_{i} g_\rho^i \tau_3^i\rho_i^v = 0.
\end{align}
The terms $\frac{\partial m_\omega}{\partial \mathrm{\sigma}}$ and $\frac{\partial m_\rho}{\partial \mathrm{\sigma}}$ originate from the dependence of the vector meson masses on the scalar field $\mathrm{\sigma}$, while $\tau_3^i$ denotes the third component of the isospin of the corresponding nucleon. The constants and parameters associated with this formalism are listed in Table~\ref{tab:parameters}. The number density, $\rho_i^v$, have different forms for protons and neutrons. For protons, the expression is given by \cite{Kumari:2020mci, Kumar:2019tiw, Kumar:2018ujk}
\begin{align}
\rho_p^v &= \frac{|q_p|B}{2\pi^2} \Bigg[ \Bigg(\sum_{\nu=0}^{\nu_{\text{max}}}
\int_{0}^{\infty} dk_\parallel  \left\{
\frac{1}{1 + e^{\beta(\tilde{E}_{k,\nu,s}^p - \mu_p^*)}}
- \frac{1}{1 + e^{\beta(\tilde{E}_{k,\nu,s}^p + \mu_p^*)}}
\right\} \Bigg)_{s=+1} \nonumber \\
&+ \Bigg(\sum_{\nu=1}^{\nu_{\text{max}}}
\int_{0}^{\infty} dk_\parallel  \left\{
\frac{1}{1 + e^{\beta(\tilde{E}_{k,\nu,s}^p - \mu_p^*)}}
- \frac{1}{1 + e^{\beta(\tilde{E}_{k,\nu,s}^p + \mu_p^*)}}
\right\} \Bigg)_{s=-1} \Bigg].
\end{align}
The number density of neutron is evaluated using 
\begin{align}
\rho_n^v &= \sum_{s=\pm 1} \int \frac{d^3k}{(2\pi)^3}  \left[
\frac{1}{1 + e^{\beta(\tilde{E}_{k,s}^n - \mu_n^*)}}
- \frac{1}{1 + e^{\beta(\tilde{E}_{k,s}^n + \mu_n^*)}}
\right].
\end{align}
Here, the scalar density associated with the nucleon $i$ is given by $\rho_i^\text{S} = \langle \bar{\psi}_i \psi_i \rangle = \frac{\partial \Omega}{\partial m_i^*}$. The scalar fields $\mathrm{\sigma}$, $\mathrm{\zeta}$, and $\mathrm{\delta}$ are determined self-consistently through a set of coupled equations, as they depend on the scalar densities $\rho_i^\text{S}$. In turn, these scalar densities are functions of the effective nucleon masses and hence of the scalar fields themselves, leading to a self-consistent solution. When the contributions from the DS to the grand potential are neglected, the scalar density of the proton is obtained as
\begin{align}
 \rho_\text{S}^{p,\text{med}} 
&= \frac{|q_p|B}{2\pi^2} \mathcal{M}_p^*\Bigg[\Bigg(\sum_{\nu=0}^{\nu_{\text{max}}}   \int_{0}^{\infty} dk_\parallel 
\frac{\bar{m}^{p}_{\nu,s}}{\tilde{E}_{k,\nu,s}^p (\bar{m}^{p}_{\nu,s}+s\kappa_pB)}  \nonumber \\
& \times \left[
\frac{1}{1 + e^{\beta(\tilde{E}_{k,\nu,s}^p - \mu_p^*)}} + \frac{1}{1 + e^{\beta(\tilde{E}_{k,\nu,s}^p + \mu_p^*)}}
\right]\Bigg)_{s=+1}+\Bigg(\sum_{\nu=1}^{\nu_{\text{max}}}   \int_{0}^{\infty} dk_\parallel 
\frac{\bar{m}^{p}_{\nu,s}}{\tilde{E}_{k,\nu,s}^p (\bar{m}^{p}_{\nu,s}+s\kappa_pB)} \nonumber \\ 
& \times  \left[
\frac{1}{1 + e^{\beta(\tilde{E}_{k,\nu,s}^p - \mu_p^*)}}
+ \frac{1}{1 + e^{\beta(\tilde{E}_{k,\nu,s}^p + \mu_p^*)}}
\right]\Bigg)_{s=-1} \Bigg], 
\end{align}
and the scalar density associated with neutrons can be expressed by
\begin{align}
 \rho_\text{S}^{n,\text{med}} 
= m_n^* \int \frac{d^3k}{(2\pi)^3} \sum_{s=\pm 1}   \frac{\bar{m}^n_{k,s}}{\tilde{E}_{k,s}^n (\bar{m}^n_{k,s}+ s \kappa_n B)}  \left[
\frac{1}{1 + e^{\beta(\tilde{E}_{k,s}^n - \mu_n^*)}}
+ \frac{1}{1 + e^{\beta(\tilde{E}_{k,s}^n + \mu_n^*)}}
\right].
\end{align}
These densities are incorporated into the coupled nonlinear field equations to evaluate the medium modifications of hadronic properties  (in particular, pions are taken for this work). 
%Since the magnetic field is incorporated only at the nucleon level in the present framework, the direct coupling of the magnetic field to pions and their constituent quarks is neglected. 
While conventional approaches often neglect the nucleon DS contribution, the present analysis includes its effects, with the magnetic field incorporated only at the nucleon level under the weak-field approximation. In this framework, a series expansion of the nucleon propagator is performed in powers of $q_i B$ and $\kappa_i B$, preserving terms up to the second order. The purely vacuum contribution to the self-energy, being comparatively small, is neglected. The inclusion of DS effects under a finite magnetic field leads to changes in the scalar densities of nucleons. As a result, an extra contribution appears in the scalar density, which is given by \cite{Mishra:2023uhx, Mukherjee:2018ebw}
\begin{align}
\rho_\text{S}^{DS,i} &= - \frac{1}{4\pi^2} \left[
\frac{(q_i B)^2}{3 m_i^*}
+ \left\{ (\kappa_i B)^2 m_i^* + (|q_i| B)(\kappa_i B) \right\}
\left( \frac{1}{2} + 2 \ln \left( \frac{m_i^*}{m_i} \right) \right)
\right].
\end{align}
\subsection{Light-Cone Quark Model (LCQM)}
\par The changes arising from isospin asymmetry, baryon density, temperature, and the influence of magnetic fields are included in the light-front framework through effective masses of quark evaluated from the CQMF model. 
%The light-cone wave functions (LCWFs) are used to calculate PDFs and EMFFs using these modified quark masses as input. The leading quark–antiquark component predominantly governs the pion, which is described as a superposition of Fock states within the light-front formalism as \cite{Luan:2024dvc} 
The pion state is described as a superposition of Fock states within the light-front formalism as \cite{Luan:2024dvc} 
\begin{equation}
|\pi\rangle = |q\bar{q}\rangle + |q\bar{q}g\rangle + \cdots, 
\end{equation}
in which the leading quark–antiquark component predominantly governs the pion structure. Therefore, we consider only the lowest Fock state. The state corresponding to the longitudinal spin projection $S_z = 0$ can be expressed in terms of the LCWF $\Psi^{\lambda_1,\lambda_2}_{\text{eff}} (x, \mathrm{k}_t)$ by
 \cite{Lepage:1980fj, Qian:2008px,Kaur:2025sgh}
\begin{equation}
|\pi \ (P^+, P_t)\rangle = \sum_{\lambda_1,\lambda_2}
\int \frac{dx\, d^2 \mathrm{k}_t}{16\pi^3 \sqrt{x(1-x)}}
\, \Psi^{\lambda_1,\lambda_2}_{\text{eff}} (x, \mathrm{k}_t)
\, |x, \mathrm{k}_t, \lambda_1, \lambda_2\rangle.
\end{equation}
Here, $x$ denotes the quark's longitudinal momentum fraction, and $\mathrm{k}_t$ represents its transverse momentum. The total momentum of the pion is given by $P = (P^+, P^-, P_t)$, while $\lambda_1$ and $\lambda_2$ are associated with the the quark and antiquark helicities, respectively. The four-momentum of the pion, along with those of the quark $q$ and antiquark $\bar{q}$ constituents, can be expressed as
\begin{equation}
\begin{aligned}
P &= \left( P^{+}, \frac{M^{*2}}{P^{+}}, \mathbf{0}_{t} \right), \\
k_q &= \left( xP^{+}, \frac{\mathrm{k}_t^{2} + m_q^{*2}}{xP^{+}}, \mathrm{k}_t \right), \\
k_{\bar{q}} &= \left( (1-x)P^{+}, \frac{\mathrm{k}_t^{2} + m_{\bar{q}}^{*2}}{(1-x)P^{+}}, -\mathrm{k}_t \right),
\end{aligned}
\end{equation}
respectively. $M^{*}$ denotes the effective invariant mass of the pion in the medium, which can be expressed in terms of the in-medium masses of its constituent quark $m_q^*$ and antiquark $m_{\bar{q}}^*$ as \cite{Xiao:2002iv}
\begin{align}
M^{*2} = \frac{\mathrm{k}_t^2 + m_q^{*2}}{x} + \frac{\mathrm{k}_t^2 + m_{\bar{q}}^{*2}}{1 - x}.
\end{align}
The total LCWF $\Psi^{\lambda_1,\lambda_2}_{\text{eff}} (x, \mathrm{k}_t)$, can be formulated as the combination of spin $\Phi_{\text{eff}}^{\lambda_1,\lambda_2}(x,\mathrm{k}_t)$ and momentum space components $\varphi_{\text{eff}}(x,\mathrm{k}_t)$ as

\begin{align}
    \Psi^{\lambda_1,\lambda_2}_{\text{eff}} (x, \mathrm{k}_t)=\Phi_{\text{eff}}^{\lambda_1,\lambda_2}(x,\mathrm{k}_t)\ \varphi_{\text{eff}}(x,\mathrm{k}_t).
\end{align}
The spin components of LCWFs are given by \cite{Luan:2024dvc,Qian:2008px}

\begin{equation}
\begin{aligned}
\Phi^{\uparrow,\uparrow}_{\text{eff}} (x,\mathrm{k}_t) 
&= -\frac{1}{\sqrt{2}} 
\frac{k_x - i k_y}{\sqrt{\mathrm{k}_t^2 + m_q^{*2}}},
\quad &
\Phi^{\uparrow,\downarrow}_{\text{eff}} (x,\mathrm{k}_t) 
&= \frac{1}{\sqrt{2}} 
\frac{m_q^*}{\sqrt{\mathrm{k}_t^2 + m_q^{*2}}}, \\[8pt]
\Phi^{\downarrow,\uparrow}_{\text{eff}} (x,\mathrm{k}_t) 
&= -\frac{1}{\sqrt{2}} 
\frac{m_q^*}{\sqrt{\mathrm{k}_t^2 + m_q^{*2}}},
\quad &
\Phi^{\downarrow,\downarrow}_{\text{eff}} (x,\mathrm{k}_t) 
&= -\frac{1}{\sqrt{2}} 
\frac{k_x + i k_y}{\sqrt{\mathrm{k}_t^2 + m_q^{*2}}}.
\end{aligned}
\label{eq:spin_wf}
\end{equation}
For charged pions $\pi^+$ and $\pi^-$, the same mass value can be assigned to both the quark and antiquark, as the in-medium mass is governed by the third component of isospin, which is the same for $u$ and $\bar{d}$, and likewise for $d$ and $\bar{u}$. Accordingly, we can take $m_q^* = m_{\bar{q}}^*$. The pion momentum-space wave function is constructed within the Brodsky-Huang-Lepage (BHL) framework and is given by \cite{Xiao:2002iv,Yu:2007hp}
\begin{equation}
\varphi_{\text{eff}}(x,\mathrm{k}_t) = \mathcal{A} \exp \left[
-\frac{1}{8\beta_\pi^2}
\left(
\frac{m_q^{*2} + \mathrm{k}_t^2}{x(1-x)}
\right)
\right],
\end{equation}
where $\mathcal{A}$ denotes the normalization constant which is evaluated for every set of masses, while $\beta_\pi$ represents the harmonic oscillator scale parameter. The momentum-space wave function follows the normalization condition
\begin{equation}
\int \frac{dx\, d^2 \mathrm{k}_t}{2(2\pi)^3} \, |\varphi_{\text{eff}}(x,\mathrm{k}_t)|^2 = 1.
\end{equation}
To evaluate the PDFs and EMFFs, the quark-quark correlator for the system of pion in vacuum state is given by \cite{Kaur:2018ewq}
\begin{equation}
H^{q}(x, \xi, q^{2}) = \frac{1}{2} \int \frac{dz^{-}}{2\pi} \, e^{i k \cdot z} 
\left\langle \pi(P_{f}) \right| 
\bar{\psi}(-z/2)\,\gamma^{+}\,\psi(z/2) 
\left| \pi(P_{i}) \right\rangle 
\Big|_{z^{+}=0,\, \mathbf{z}_{t}=0},
\end{equation}
where $-q^{2} = Q^{2}$ is the squared four-momentum transfer, expressed in units of GeV$^{2}$, and $P_f$ and $P_i$ denote the final and initial momenta of the pion, respectively. For the case of $P_f=P_i$, the function $H^{q}(x, \xi = 0, q^{2})$ reduces to the ordainary unpolarized PDF $f^{q}(x)$. PDFs characterize how the momentum of a hadron is distributed among its constituent partons. They provide the likelihood of finding a parton with a fraction $x$ of the momentum of the hadron. By incorporating $m_q^*$, the unpolarized valence quark PDF $f_{\text{eff}}^{q}(x)$ is expressed as \cite{Maji:2016yqo}
\begin{equation}
f_{\text{eff}}^{q}(x) = \int \frac{d^{2}\mathrm{k}_t}{16\pi^{3}} \sum_{\lambda_1,\lambda_2} \left| \Psi_{\text{eff}}^{\lambda_1,\lambda_2}(x,\mathrm{k}_t) \right|^{2}.
\label{eq:pdf}
\end{equation}
%
%The explicit nature of LCWFs further simplifies this expression to
%
%\begin{equation}
%f_{\pi}^{q}(x) = \int \frac{d^{2}\mathrm{k}_\perp}{16\pi^{3}}\left| \varphi_\pi(x,\mathrm{k}_\perp) \right|^{2}.
%\label{eq:pdf}
%\end{equation}

The unpolarized PDFs satisfy the normalization (sum rule) conditions,
\begin{align}
&\int dx f_{\text{eff}}^{q}(x) = \int dx  f_{\text{eff}}^{\bar{q}}(x) = 1, 
\end{align}
where $f_{\text{eff}}^{\bar{q}}(x)$ denotes the antiquark distribution function of the pion.
\par EMFFs of the quark, $F^q_{\text{eff}}(Q^2)$ describe the distribution of electric charge and current inside a hadron. They give understanding of the internal structure and spatial extent of the particle. It can be defined from GPDs, when the skewness parameter ($\xi$) is set to zero. 
%GPDs can be evaluated from the quark--quark correlator, which is given by \cite{Kaur:2018ewq}
%\begin{equation}
%H_{\pi}^{q}(x, \xi = 0, q^{2}) = \frac{1}{2} \int \frac{dz^{-}}{2\pi} \, e^{i k \cdot z} 
%\left\langle \pi(P_{f}) \right| 
%\bar{\psi}(-z/2)\,\gamma^{+}\,\psi(z/2) 
%\left| \pi(P_{i}) \right\rangle 
%\Big|_{z^{+}=0,\, \mathbf{z}_{\perp}=0},
%\end{equation}
%
%where $-q^{2} = Q^{2}$ is the squared four-momentum transfer, expressed in units of GeV$^{2}$, and $P_f$ and $P_i$ denote the final and initial momenta of the pion, respectively. 
In terms of the overlap form of LCWFs, EMFF is defined as \cite{Kaur:2018ewq}
\begin{equation}
F_{\text{eff}}^{q}(Q^{2}) = \int \frac{dx \, d^{2}\mathrm{k}_t}{16\pi^{3}} 
\sum_{\lambda_{1},\lambda_{2}} 
\Psi_{\text{eff}}^{*\lambda_{1},\lambda_{2}}(x,\mathrm{k}_t^{\prime\prime}) \,
\Psi_{\text{eff}}^{\lambda_{1},\lambda_{2}}(x,\mathrm{k}_t^{\prime}),
\end{equation}
where the transverse momenta of the constituent valence quark in the final and initial states are given by $\mathrm{k}_t^{\prime\prime} = \mathrm{k}_t - (1 - x)\mathrm{q}_t/2$ and $\mathrm{k}_t^\prime = \mathrm{k}_t+ (1 - x)\mathrm{q}_t/2$, respectively. Inserting the spin wave function from Eq.~(\ref{eq:spin_wf}) into the overlap representation, further simplifies the expression to
\begin{equation}
F_{\text{eff}}^{q}(Q^{2}) = \int \frac{dx \, d^{2}\mathrm{k}_t}{16\pi^{3}} 
\left[ 
\mathrm{k}_t^{2}+m_{q}^{*2} - \frac{(1-x)^{2}Q^{2}}{4} 
\right]
\frac{\varphi_{\text{eff}} (x,\mathrm{k}_t^{\prime\prime}) \, \varphi_{\text{eff}}(x,\mathrm{k}_t^{\prime})}
{\sqrt{\mathrm{k}_t^{\prime\prime2} + m_q^{*2}} \, \sqrt{\mathrm{k}_t^{\prime2} + m_q^{*2}}}.
\label{eq:emff}
\end{equation}
\par The PDFs and EMFFs of the corresponding antiquark can be obtained by interchanging $m_q^*$ with the in-medium antiquark mass $m_{\bar{q}}^*$ in Eqs.~(\ref{eq:pdf}) and (\ref{eq:emff}).
\begin{table}[h]
\centering
\renewcommand{\arraystretch}{1.5}
\begin{tabular}{|c|c|c|c|c|c|c|}
\hline
%
% Row 1
~~~~~~~$k_0$~~~~~~~  & ~~$m_\mathrm{\omega}\ (\mathrm{MeV})$~~ & ~~$m_\mathrm{\rho}\ (\mathrm{MeV})$~~ & ~~~~~~$g^{u,d}_{\mathrm{\sigma}}$~~~~~~ & ~~~~~~$g^{s}_{\mathrm{\sigma}}$ ~~~~~~& $~~~~~~g^{p,n}_{\mathrm{\sigma}}$~~~~~~~&~~~~~~~$g_4$ ~~~~~~~\\
% Row 2
4.94  & 783 & 783& 2.72 & 0  &6.64 & 37.5\\
\hline
%
% Row 3
$k_1$  & $m_\pi\ (\mathrm{MeV})$  &  $ \mathrm{\sigma_0\ (MeV)}$ & $g^{u,d}_{\mathrm{\zeta}}$ & $g^{s}_{\mathrm{\zeta}}$ &$g^{p,n}_{\mathrm{\zeta}}$& $d$\\
% Row 4
2.12  & 139 & -93& 0 & 3.847   &0& 0.182\\
\hline
%
% Row 5
$k_2$  & $m_K\ (\mathrm{MeV})$  & $ \mathrm{\zeta_0\ (MeV)}$& $g^{u,d}_{\mathrm{\delta}}$ & $g^{s}_{\mathrm{\delta}}$   & $g^{p,n}_{\mathrm{\delta}}$&$\kappa_p$($\mu_N$) \\
% Row 6
-10.16  & 496 & -96.87& 2.72 & 0  &2.72& 1.793\\
\hline
%
% Row 7
$k_3$  & $f_\pi\ (\mathrm{MeV})$ & $\rho_0$ \ ($fm^{-3}$)& $g^{u,d}_{\mathrm{\rho}}$ & $g^{s}_{\mathrm{\rho}}$   & $g^{p,n}_{\mathrm{\rho}}$&$\kappa_n$($\mu_N$) \\
% Row 8
-5.38  & 93 & 0.16 & 3.23  & 0 &8.89& -1.913\\
\hline
%
% Row 9
$k_4$  & $f_K\ (\mathrm{MeV})$ & $ \mathrm{\chi_0\ (MeV)}$& $g^{u,d}_{\mathrm{\omega}}$ & $g_{\mathrm{\omega}}^s$ & $g_{\mathrm{\omega}}^{p,n}$&$\beta_\pi$\\
% Row 10
-0.06  & 115 & 254.38 & 3.23 & 0 &9.69& 0.41\\
\hline
\end{tabular}
\caption{The table presents the coupling constants and parameters used in this study.}
\label{tab:parameters}
\end{table}%
\section{Results and Discussion}
\label{sec:Results and Discussion}
\par In the present work, the combined effects of temperature ($T\ (\mathrm{MeV})$), baryon density ($\mathrm{\rho_B/\rho_0}$), isospin asymmetry ($\mathrm{\eta}$) and an external magnetic field ($\mathrm{eB/m_\pi^2}$) are found to significantly in altering the characteristics of the medium, primarily through changes in the scalar and vector fields. These fields play a crucial role in determining the in-medium behavior of hadrons, as they directly affect the effective masses of the constituent quarks $m_q^*$. Variations in $m_q^*$ further influence the internal structure of the pion, which is reflected in its PDFs and EMFFs. Thus, the analysis provides a detailed understanding of how medium modifications, mediated by scalar and vector fields, govern the properties of the pion under extreme conditions.
\subsection{Scalar fields in a hot nuclear medium under a magnetic field}
\par
\begin{figure}[t]
    \centering
    % Column 1
    
            \includegraphics[width=\linewidth,clip,trim=1.4cm 10cm 1.5cm 9cm]{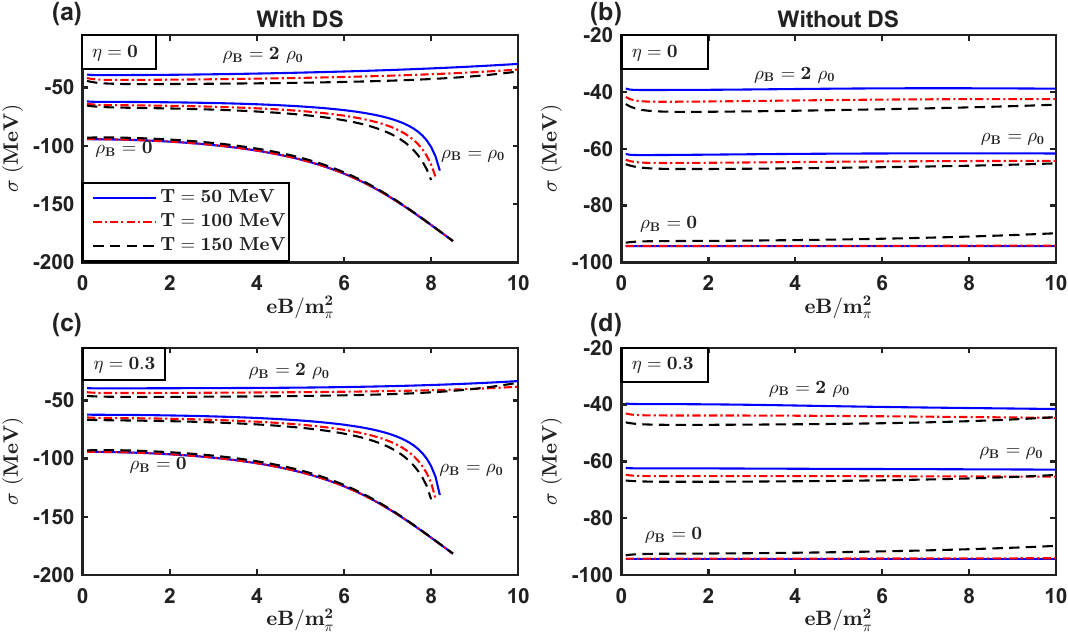}

    \caption{Dependence of the scalar field $\mathrm{\sigma}$ on the magnetic field 
$\mathrm{{eB}/{m_{\pi}^2}}$ for baryon densities $\mathrm{\rho_B = 0}$, $\mathrm{\rho_0}$, 
and $\mathrm{2\rho_0}$. The findings are displayed for both symmetric ($\mathrm{\eta = 0}$) and 
asymmetric ($\mathrm{\eta = 0.3}$) nuclear matter at temperatures $\mathrm{T = 50}$, $\mathrm{100}$, and 
$\mathrm{150}$ MeV. The plots compare scenarios with and without the inclusion of DS effects.}
\label{fig:sigma}   
\end{figure}

\begin{figure}[htbp]
    \centering
    % Column 1
   \includegraphics[width=\linewidth,clip,trim=1.4cm 10cm 1.5cm 9cm]{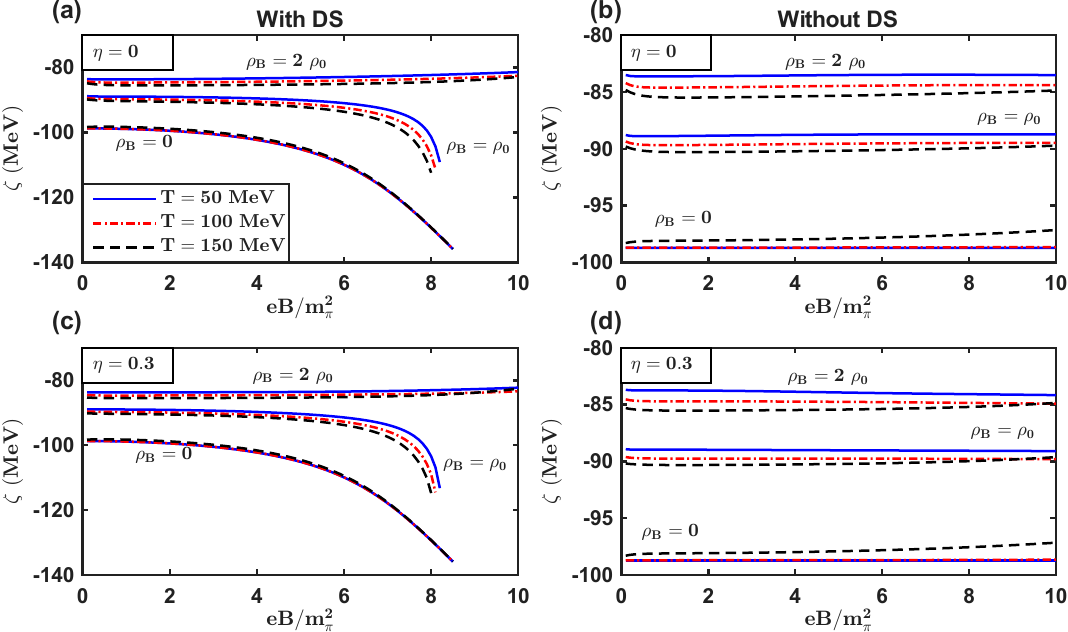}
    \caption{Dependence of the $\mathrm{\zeta}$ field on the magnetic field 
$\mathrm{{eB}/{m_{\pi}^2}}$ under the same conditions as described in Fig.~\ref{fig:sigma}.}
\label{fig:zeta} 
\end{figure}

\begin{figure}[htbp]
    \centering
    % Column 1
   \includegraphics[width=\linewidth,clip,trim=1.4cm 10cm 1.5cm 9cm]{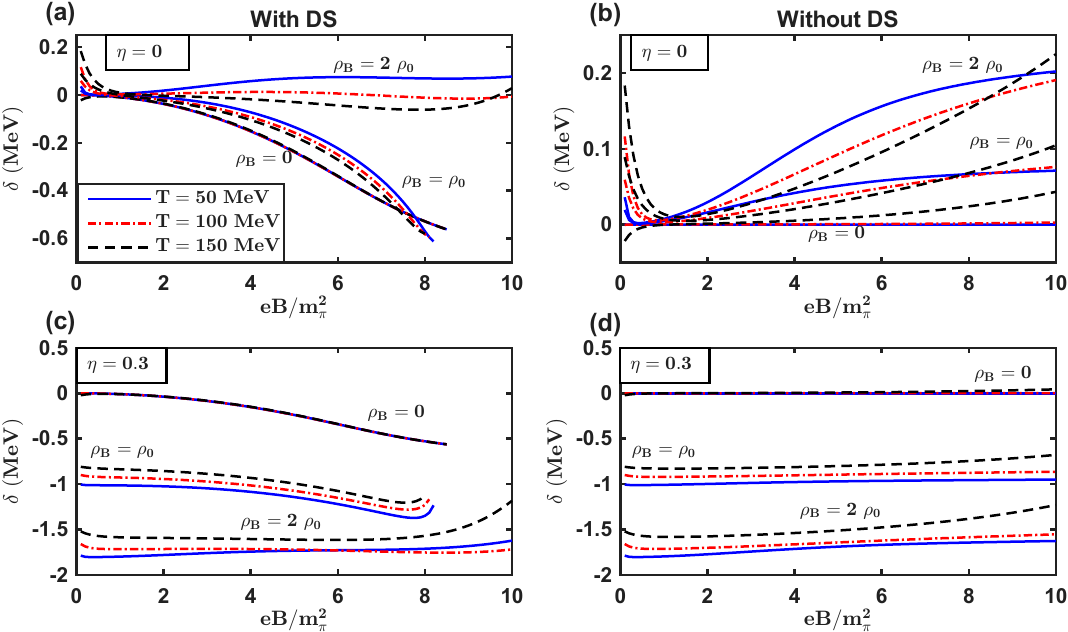}
    \caption{Dependence of the $\mathrm{\delta}$ field on the magnetic field 
$\mathrm{{eB}/{m_{\pi}^2}}$ under the same conditions as described in Fig.~\ref{fig:sigma}.}
\label{fig:delta}
\end{figure}

 Figures~\ref{fig:sigma}, \ref{fig:zeta} and \ref{fig:delta} illustrate the variation of respective scalar fields $\mathrm{\sigma}$, $\mathrm{\zeta}$, and $\mathrm{\delta}$ with magnetic field strength at  $\mathrm{\rho_B} = 0$, $\rho_0$, and $2\rho_0$, for finite temperatures $\mathrm{T} = 50, 100,$ and $150$~MeV. Results are presented for both symmetric ($\mathrm{\eta} = 0$) and asymmetric ($\mathrm{\eta} = 0.3$) nuclear matter, in row 1 and 2, respectively. Each case is analyzed with DS effects included (left panels) and compared to results without them (right panels). Comparing the magnitudes of the fields $\mathrm{\sigma}$ and $\mathrm{\zeta}$ with and without the DS, a pronounced variation is observed as the function of the magnetic field, when DS is included. In its absence the fields remain nearly constant, with only negligible changes in some cases, as evident in Figs.~\ref{fig:sigma} and \ref{fig:zeta}. Upon incorporating nucleon AMMs, the $\mathrm{\sigma}$ and $\mathrm{\zeta}$ fields exhibit an increasing trend in magnitude with the magnetic field at $\mathrm{\rho_B} = 0$ and $\mathrm{\rho_B} = \mathrm{\rho_0}$, consistent with magnetic catalysis (MC). In contrast, at higher density ($\mathrm{\rho_B} = 2\rho_0$), the opposite trend emerges as the magnitudes of the $\mathrm{\sigma}$ and $\mathrm{\zeta}$ fields decrease with increasing magnetic field, reflecting the phenomenon of inverse magnetic catalysis (IMC).
\par At $\mathrm{eB} = 4\,\mathrm{m_{\pi}^2}$, the scalar field $\mathrm{\sigma}$ in symmetric (asymmetric) nuclear matter at saturation density $\mathrm{\rho_B} = \rho_0$ takes values of $-64.302\,(-65.228)$, $-67.782\,(-68.391)$, and $-70.523\,(-70.721)$ in the units of $\mathrm{MeV}$ for $\mathrm{T} = 50, 100,$ and $150$~MeV, respectively. Increasing the magnetic field to $\mathrm{eB} = 8\,\mathrm{m_{\pi}^2}$ modifies these values to $-100.98\,(-106.55)$, $-115.98\,(-120.81)$, and $-129.31\,(-134.98)$~MeV. The corresponding percentage changes are $57.04\%\,(63.35\%)$, $71.10\%\,(76.65\%)$, and $83.36\%$ $(90.86\%)$ at $\mathrm{T} = 50, 100,$ and $150$~MeV. These results highlight a clear temperature dependence as the relative variation grows with increasing $\mathrm{T}$, reflecting a stronger sensitivity of the medium to magnetic fields at higher thermal excitation. High temperature domain open additional phase-space states, thereby amplifying contributions from the DS and Landau quantization, which enhance the magnetic field response of $\mathrm{\sigma}$ field. A similar trend is observed when the system evolves from symmetric to asymmetric nuclear matter, indicating that a larger $\mathrm{\eta}$ further strengthens MC effects.
\par At higher density, $\mathrm{\rho_B} = 2\rho_0$, the scalar fields exhibit a distinct behavior. The magnitude of the $\mathrm{\sigma}$ field at $\mathrm{\eta} = 0$ ($\mathrm{\eta} = 0.3$) for magnetic field $\mathrm{eB} = 4\,\mathrm{m_{\pi}^2}$ is $-37.947\,(-39.419)$, $-42.32\,(-43.335)$, and $-46.32\,(-46.687)$~MeV for $\mathrm{T} = 50, 100,$ and $150$~MeV, respectively. As the magnetic field strength increases to $\mathrm{eB} = 8\,\mathrm{m_{\pi}^2}$, the values change to $-33.539\,(-36.922)$, $-38.515\,(-41.379)$, and $-42.62\,(-43.211)$ $\mathrm{MeV}$, respectively. The percentage changes are $11.62\%\,(6.33\%)$, $8.99\%\,(4.51\%)$, and $7.99\%$, $(7.44\%)$ for the respective temperatures. Unlike the earlier case, the percentage change decreases with rising temperature, showing that thermal effects partially weaken IMC at high density. As temperature increases, thermal excitations reduce the role of the dense FS, thereby diminishing the suppression caused by magnetic fields and resulting in smaller percentage changes. The percentage drop is reduced in the asymmetric medium, indicating a weakening of IMC with asymmetry. A slight increase in the percentage change at $\mathrm{T} = 150$~MeV for the asymmetric case suggests a competition between thermal effects and magnetic field induced modifications. This shows that although IMC remains, its strength is influenced by the combined effects of density, temperature, and isospin asymmetry. A similar trend is observed at higher magnetic fields for both symmetric and asymmetric media. For $\mathrm{eB} = 10\,\mathrm{m_{\pi}^2}$, the magnitude of $\mathrm{\sigma}$ is calculated as $-29.663\,(-33.54)$, $-34.476\,(-38.366)$, and $-36.141\,(-34.902)$~MeV at $\mathrm{T} = 50, 100,$ and $150$~MeV, respectively. The percentage changes relative to $\mathrm{eB} = 8\,\mathrm{m_{\pi}^2}$ are $11.56\%\,(9.16\%)$, $10.49\%\,(7.281\%)$, and $15.201\%\,(19.228\%)$. It is observed that the percentage change increases compared to the lower magnetic field case. Although IMC reduces the scalar field strengths with increasing magnetic field, the rate of suppression becomes larger at stronger magnetic field. In other words, the magnetic-field-induced modifications to the scalar densities, through Landau quantization and Dirac sea contributions, grow nonlinearly with $\mathrm{B}$, leading to a stronger relative change as magnetic field strengthens from $\mathrm{eB} = 8$ to $10\,\mathrm{m_{\pi}^2}$. The same trend is visible for the scalar field $\mathrm{\zeta}$ in Fig. \ref{fig:zeta}.
\par Figure~\ref{fig:delta} illustrates how the $\mathrm{\delta}$ field varies with magnetic field under different conditions. A zero-field value was expected at $\mathrm{\rho_B} = 0$ in the symmetric medium. However, a small non-zero value appears when DS and AMM effects are included. This arises because the magnetic field influences protons and neutrons differently due to their unequal electric charges and magnetic moments, leading to unequal scalar densities even in the absence of net baryon density. Consequently, an effective isospin asymmetry is dynamically generated in the system, which induces a finite $\mathrm{\delta}$ field despite the nominally symmetric conditions.The inclusion of DS contributions does not lead to a significant change in the magnitude of the $\delta$ field when compared to the $\sigma$ and $\zeta$ fields. However, it noticeably affects the overall trend. At lower baryon densities, the $\delta$ field decreases with increasing magnetic field strength ($eB/m_\pi^2$), whereas at higher density ($\rho_B = 2\rho_0$), the trend reverses, which can be attributed to the inverse magnetic catalysis (IMC) effect. Furthermore, the $\delta$ field exhibits a comparatively stronger sensitivity to the magnetic field when DS contributions are included. Finite temperature effects tend to smooth out these variations, leading to a reduction in the overall magnitude of the field. The differences observed between symmetric and asymmetric matter highlight the role of isospin asymmetry ($\eta$) in determining the response of the $\delta$ field in a magnetized medium. 
\subsection{In-Medium Modification of Quark Masses in Hot Magnetized Nuclear Matter}
\begin{table*}
\centering
\renewcommand{\arraystretch}{1.5}
\fontsize{10.4}{12}\selectfont
\begin{tabular}{|c|c|c|c|c|c|c|c|c|}
\hline
\hline
% Header Row 1
\makecell{~Density~ \\ ($\mathrm{\rho_B/\rho_0}$)} & \makecell{Isospin \\ ~Asymmetry~ \\ $\mathrm{\eta}$} & \makecell{~Temperature~ \\ $\mathrm{T}$ \\($\mathrm{MeV}$)} 
& \multicolumn{6}{c|}{\makecell{Modified $\mathrm{u}$ Quark Mass $\mathrm{m_u^*}$  \\ ($\mathrm{MeV}$)}} \\
\cline{4-9}

% Header Row 2
& & 
& ~~~~$\mathrm{eB=0}~~~$ 
&~~~~ $\mathrm{2m_\pi^2}$~~~ 
& ~~~~$\mathrm{4m_\pi^2}$ ~~~
&~~~ $\mathrm{6m_\pi^2}$ ~~
& ~~~$\mathrm{8m_\pi^2}$ ~~
&~~ $\mathrm{10m_\pi^2}$~~ \\
\hline

% Sample Rows den0
\multirow{6}{*}{$0$} &  0  & 50  & \makecell{256.45 \\ (256.45)} & \makecell{ 262.24\\ (256.45)} & \makecell{ 283.42\\ (256.45)} & \makecell{ 337.44\\ (256.45)} & \makecell{ 465.05\\ (256.45)} & \makecell{ -\\ (256.45)} \\
\cline{3-9}

% & 0   & 100 & \makecell{256.34\\ (256.34)} & \makecell{ 262.14\\ (256.33)} & \makecell{283.32 \\ (256.3)} & \makecell{337.38 \\ (256.23)} & \makecell{457.64 \\ (256.11)} & \makecell{ -\\ (255.93)} \\
%\cline{3-9}

 &     & 150  & \makecell{251.67\\ (251.67)} & \makecell{257.4 \\ (251.44)} & \makecell{278.68 \\ (250.71)} & \makecell{334.29 \\ (249.38)} & \makecell{457.16 \\ (247.28)} & \makecell{ -\\ (244.08)} \\
\cline{2-9}

 &  0.3  & 50  & \makecell{256.45\\ (256.45)} & \makecell{262.24 \\ (256.45)} & \makecell{ 283.42\\ (256.45)} & \makecell{337.44 \\ (256.45)} & \makecell{ 457.64\\ (256.45)} & \makecell{ -\\ (256.45)} \\
\cline{3-9}

% & 0.3   & 100 & \makecell{256.34\\ (256.34)} & \makecell{ 262.14\\ (256.33)} & \makecell{ 283.32\\ (256.3)} & \makecell{337.38 \\ (256.23)} & \makecell{ 457.64\\ (256.11)} & \makecell{ -\\ (255.93)} \\
%\cline{3-9}

 &      & 150  & \makecell{251.67\\ (251.67)} & \makecell{257.4 \\ (251.44)} & \makecell{278.68 \\ (250.71)} & \makecell{334.29 \\ (249.38)} & \makecell{ 457.16\\ (247.28)} &\makecell{ -\\ (244.08)}  \\

\hline

% Sample Rows den1
\multirow{6}{*}{$1$}& 0   & 50  & \makecell{169.30 \\ (169.30)} & \makecell{170.36 \\ (168.79)} & \makecell{175.03 \\ (168.08)} & \makecell{ 188.96\\ (167.65)} & \makecell{275.47 \\ (167.54)} & \makecell{ -\\ (167.64)} \\
\cline{3-9}

% &  0  & 100 & \makecell{176.96\\ (176.96)} & \makecell{178.5 \\ (176.63)} & \makecell{ 184.52\\ (175.92)} & \makecell{201.94 \\ (175.29)} & \makecell{316.32 \\ (174.9)} & \makecell{ -\\ (174.72)} \\
%\cline{3-9}

 &     & 150  & \makecell{182.87\\ (182.87)} & \makecell{184.76 \\ (182.58)} & \makecell{ 192\\ (181.9)} & \makecell{212.7 \\ (180.84)} & \makecell{352.57 \\ (179.36)} &\makecell{ -\\ (177.06)}  \\
\cline{2-9}

 &  0.3  & 50  & \makecell{171.25\\ (171.25)} & \makecell{172.87 \\ (171.3)} & \makecell{178.92 \\ (171.57)} & \makecell{ 194.84\\ (171.83)} & \makecell{291.68 \\ (172.69)} & \makecell{ -\\ (172.38)} \\
\cline{3-9}

 %& 0.3   & 100 & \makecell{178.64\\ (178.64)} & \makecell{180.5 \\ (178.6)} & \makecell{187.43 \\ (178.56)} & \makecell{206.58 \\ (178.64)} & \makecell{330.32 \\ (178.81)} & \makecell{ -\\ (179.03)} \\
%\cline{3-9}

 &      & 150  & \makecell{183.98 \\ (183.98)} & \makecell{ 186.03\\ (183.82)} & \makecell{193.66 \\ (183.43)} & \makecell{ 215.43\\ (182.38)} & \makecell{ 368.72\\ (180.55)} & \makecell{ -\\ (177.08)} \\

\hline

% Sample Rows den2
\multirow{6}{*}{$2$} & 0   & 50   & \makecell{106.91\\ (106.91)} & \makecell{106.08 \\ (106.7)} & \makecell{ 103.15\\ (105.86)} & \makecell{98.309 \\ (105.15)} & \makecell{ 91.15\\ (105.03)} & \makecell{80.59 \\ (105.38)} \\
\cline{3-9}

 %& 0   & 100 & \makecell{118.33\\ (118.33)} & \makecell{117.5 \\ (117.89)} & \makecell{115.11 \\ (116.89)} & \makecell{111.16 \\ (115.97)} & \makecell{104.79 \\ (115.47)} & \makecell{115.39 \\ (115.39)} \\
%\cline{3-9}

 &     & 150  & \makecell{128.19\\ (128.19)} & \makecell{127.63 \\ (127.76)} & \makecell{126.03 \\ (126.73)} & \makecell{122.84 \\ (125.25)} & \makecell{ 116.03\\ (123.36)} & \makecell{ 98.28\\ (120.65)} \\
\cline{2-9}

 &  0.3  & 50  & \makecell{110.73\\ (110.73)} & \makecell{ 110.38\\ (110.96)} & \makecell{109.61 \\ (111.99)} & \makecell{ 107.59\\ (113.3)} & \makecell{102.77 \\ (114.32)} & \makecell{ 93.45\\ (115.23)} \\
\cline{3-9}

 %& 0.3   & 100 & \makecell{121.55\\ (121.55)} & \makecell{121.19 \\ (121.54)} & \makecell{120.22 \\ (121.7)} & \makecell{118.48 \\ (122.18)} & \makecell{114.95 \\ (122.87)} & \makecell{ 106.71\\ (123.6)} \\
%\cline{3-9}

 &      & 150  & \makecell{130.50\\ (130.50)} & \makecell{130.19 \\ (130.3)} & \makecell{129.19 \\ (129.75)} & \makecell{126.71 \\ (128.66)} & \makecell{119.69 \\ (126.54)} & \makecell{96.56\\ (122.38)} \\

\hline
\hline

\end{tabular}

\caption{Modified mass of the $\mathrm{u}$ quark ($m_u^*$) in a magnetized nuclear medium for different baryon densities ($\mathrm{\rho_B = 0}$, $\mathrm{\rho_0}$, and $\mathrm{2\rho_0}$) at temperatures $\mathrm{T = 50}$, and $\mathrm{150}$ $\mathrm{MeV}$. The results are shown for isospin asymmetry parameters $\mathrm{\eta = 0}$ and $\mathrm{0.3}$. The values given in parentheses correspond to the case where the Dirac sea contribution is excluded.}
\label{tab:umass}

\end{table*}
\begin{table*}
\centering
\renewcommand{\arraystretch}{1.5}
\fontsize{10.4}{12}\selectfont
\begin{tabular}{|c|c|c|c|c|c|c|c|c|}
\hline
\hline
% Header Row 1
\makecell{~Density~ \\ ($\mathrm{\rho_B/\rho_0}$)} & \makecell{Isospin \\ ~Asymmetry~ \\ $\mathrm{\eta}$} & \makecell{~Temperature~ \\ $\mathrm{T}$ \\($\mathrm{MeV}$)} 
& \multicolumn{6}{c|}{\makecell{Modified $\mathrm{d}$ Quark Mass $\mathrm{m_d^*}$  \\ ($\mathrm{MeV}$)}} \\
\cline{4-9}

% Header Row 2
& & 
& ~~~~$\mathrm{eB=0}~~~$ 
&~~~~ $\mathrm{2m_\pi^2}$~~~ 
& ~~~~$\mathrm{4m_\pi^2}$ ~~~
&~~~ $\mathrm{6m_\pi^2}$ ~~
& ~~~$\mathrm{8m_\pi^2}$ ~~
&~~ $\mathrm{10m_\pi^2}$~~ \\
\hline

% Sample Rows den0
\multirow{6}{*}{$0$} &  0  & 50  & \makecell{256.45 \\ (256.45)} & \makecell{ 262.14\\ (256.45)} & \makecell{ 283.01\\ (256.45)} & \makecell{ 336.51\\ (256.45)} & \makecell{ 463.59\\ (256.45)} & \makecell{ -\\ (256.45)} \\
\cline{3-9}

% & 0   & 100 & \makecell{256.34\\ (256.34)} & \makecell{ 262.04\\ (256.33)} & \makecell{282.91 \\ (256.3)} & \makecell{336.46 \\ (256.23)} & \makecell{456.2 \\ (256.11)} & \makecell{ -\\ (255.93)} \\
%\cline{3-9}

 &     & 150  & \makecell{251.67\\ (251.67)} & \makecell{257.3 \\ (251.45)} & \makecell{278.28 \\ (250.72)} & \makecell{333.37 \\ (249.42)} & \makecell{455.73 \\ (247.35)} & \makecell{ -\\ (244.19)} \\
\cline{2-9}

 &  0.3  & 50  & \makecell{256.45\\ (256.45)} & \makecell{262.14 \\ (256.45)} & \makecell{ 283.01\\ (256.45)} & \makecell{336.51 \\ (256.45)} & \makecell{ 456.2\\ (256.45)} & \makecell{ -\\ (256.45)} \\
\cline{3-9}

 %& 0.3   & 100 & \makecell{256.34\\ (256.34)} & \makecell{ 262.04\\ (256.33)} & \makecell{ 282.91\\ (256.3)} & \makecell{336.46 \\ (256.23)} & \makecell{ 456.2\\ (256.11)} & \makecell{ -\\ (255.93)} \\
%\cline{3-9}

 &      & 150  & \makecell{251.67\\ (251.67)} & \makecell{257.3 \\ (251.45)} & \makecell{278.28 \\ (250.72)} & \makecell{333.37 \\ (249.42)} & \makecell{ 455.73\\ (247.35)} &\makecell{ -\\ (244.19)}  \\

\hline

% Sample Rows den1
\multirow{6}{*}{$1$}&  0  & 50  & \makecell{169.30 \\ (169.30)} & \makecell{170.33 \\ (168.84)} & \makecell{174.83 \\ (168.19)} & \makecell{ 188.35\\ (167.81)} & \makecell{273.92 \\ (167.72)} & \makecell{ -\\ (167.83)} \\
\cline{3-9}

 %& 0   & 100 & \makecell{176.96\\ (176.96)} & \makecell{178.45 \\ (176.65)} & \makecell{ 184.27\\ (176)} & \makecell{201.26 \\ (175.43)} & \makecell{314.73 \\ (175.07)} & \makecell{ -\\ (174.93)} \\
%\cline{3-9}

 &     & 150  & \makecell{182.87\\ (182.87)} & \makecell{184.69 \\ (182.59)} & \makecell{ 191.71\\ (181.95)} & \makecell{211.97 \\ (180.95)} & \makecell{350.99 \\ (179.54)} &\makecell{ -\\ (177.34)}  \\
\cline{2-9}

 &  0.3  & 50  & \makecell{168.49\\ (168.49)} & \makecell{170.09 \\ (168.59)} & \makecell{175.97 \\ (168.91)} & \makecell{ 191.52\\ (169.21)} & \makecell{288.01 \\ (169.21)} & \makecell{ -\\ (169.8)} \\
\cline{3-9}

% & 0.3   & 100 & \makecell{176.13\\ (176.12)} & \makecell{177.94 \\ (176.1)} & \makecell{184.67 \\ (176.11)} & \makecell{203.45 \\ (176.22)} & \makecell{326.98 \\ (176.43)} & \makecell{ -\\ (176.68)} \\
%\cline{3-9}

 &      & 150  & \makecell{181.72 \\ (181.72)} & \makecell{ 183.7\\ (181.57)} & \makecell{191.12\\ (181.13)} & \makecell{ 212.5\\ (180.24)} & \makecell{ 365.68\\ (178.52)} & \makecell{ -\\ (175.23)} \\

\hline

% Sample Rows den2
\multirow{6}{*}{$2$} &  0  & 50   & \makecell{106.91\\ (106.91)} & \makecell{106.13 \\ (106.78)} & \makecell{ 103.31\\ (106.13)} & \makecell{98.512 \\ (105.58)} & \makecell{ 91.33\\ (105.54)} & \makecell{80.8 \\ (105.93)} \\
\cline{3-9}

 %& 0   & 100 & \makecell{118.33\\ (118.33)} & \makecell{117.52 \\ (117.94)} & \makecell{115.15 \\ (117.07)} & \makecell{111.18 \\ (116.29)} & \makecell{104.76 \\ (115.9)} & \makecell{93.78 \\ (115.91)} \\
%\cline{3-9}

 &     & 150  & \makecell{128.19\\ (128.19)} & \makecell{127.62 \\ (127.79)} & \makecell{125.98 \\ (126.86)} & \makecell{122.72 \\ (125.51)} & \makecell{ 115.86\\ (123.77)} & \makecell{ 98.36\\ (121.26)} \\
\cline{2-9}

 &  0.3  & 50  & \makecell{105.82\\ (105.82)} & \makecell{ 105.54\\ (106.14)} & \makecell{104.87 \\ (107.33)} & \makecell{ 102.9\\ (108.76)} & \makecell{98.12 \\ (109.85)} & \makecell{ 89.04\\ (110.81)} \\
\cline{3-9}

% & 0.3   & 100 & \makecell{116.87\\ (116.87)} & \makecell{116.53 \\ (116.91)} & \makecell{120.22 \\ (121.7)} & \makecell{118.48 \\ (122.18)} & \makecell{114.95 \\ (122.87)} & \makecell{ 106.71\\ (123.6)} \\
%\cline{3-9}

 &      & 150  & \makecell{130.50\\ (130.50)} & \makecell{130.19 \\ (130.3)} & \makecell{129.19 \\ (129.75)} & \makecell{126.71 \\ (128.66)} & \makecell{119.69 \\ (126.54)} & \makecell{96.56\\ (122.38)} \\

\hline
\hline
\end{tabular}

\caption{Dependence of $m_d^*$ on magnetic field under identical conditions to Table~\ref{tab:umass}.}
\label{tab:dmass}

\end{table*}

\par Since the $m_q^*$ depends on the scalar fields via Eq.~(\ref{eq:mod_mass}), the combined variation of these fields is directly reflected in the quark masses. As discussed earlier, the $\mathrm{\sigma}$ and $\mathrm{\zeta}$ fields have much larger magnitudes compared to the $\mathrm{\delta}$ field. As a result, the influence of the $\mathrm{\delta}$ field on the quark masses is relatively weaker.
Additionally, because of the negative sign appearing in Eq.~(\ref{eq:mod_mass}), the variation of the quark masses exhibits a trend opposite to that of the $\mathrm{\sigma}$ and $\mathrm{\zeta}$ fields. The in-medium effective masses of $u$ and $d$ quarks are presented in Tables~\ref{tab:umass} and \ref{tab:dmass}, respectively, for selected values of magnetic field at $\mathrm{\rho_B = 0}$, $\mathrm{\rho_0}$, and $\mathrm{2\rho_0}$. The magnitudes are evaluated at $\mathrm{T = 50}$ and $\mathrm{150\ MeV}$, and $\mathrm{\eta = 0}$ and $\mathrm{\eta = 0.3}$, both with and without the inclusion of DS contributions.
\par From Tables~\ref{tab:umass} and \ref{tab:dmass}, it is evident that the effective quark mass $m_q^*$ undergoes a significant variation with the magnetic field $\mathrm{eB/m_\pi^2}$ when DS is included. This behavior arises due to vacuum polarization effects, which modify the scalar condensates through magnetic field interactions, thereby influencing the effective quark masses. In contrast, when the DS is neglected, $m_q^*$ remains nearly independent of the $\mathrm{eB/m_\pi^2}$, following the response of the scalar fields $\mathrm{\sigma}$ and $\mathrm{\zeta}$, which show minimal variation with $\mathrm{eB/m_\pi^2}$ in this case. As a result, when DS contributions are neglected, the magnetic field does not induce any significant change in the quark mass, particularly at lower baryon densities where medium effects are comparatively weak.
\par An increase in $m_q^*$ is observed at low density ($\mathrm{\rho_B=0}$ and $\mathrm{\rho_0}$), whereas it decreases at higher density $\mathrm{\rho_B = 2\rho_0}$. This contrasting behavior can be associated with the density dependence of the scalar fields, where at higher densities the stronger reduction of $\mathrm{\sigma}$ and $\mathrm{\zeta}$ fields leads to a decrease in the effective mass of quark, pointing to partial restoration of chiral symmetry. Compared to other factors, the dependence of $m_q^*$ on $\mathrm{\eta}$ is relatively weak, since its effect enters mainly through the $\mathrm{\delta}$ field, which has a smaller magnitude compared to the other scalar fields. However, a more noticeable impact is observed at higher $\mathrm{\rho_B}$ due to enhanced medium effects. A similar trend is observed for temperature, whose influence depends on both $\mathrm{\rho_B}$ and $\mathrm{\eta}$, as thermal effects modify the scalar fields differently under varying density and isospin conditions. It can thus be concluded that the modification of the quark mass with magnetic field is predominantly controlled by baryon density, whereas temperature and isospin asymmetry contribute only moderate corrections. The values obtained for $m_q^*$ at zero magnetic field and $T = 100\,\mathrm{MeV}$ are consistent with those reported in Ref.~\cite{Puhan:2024xdq}.
\subsection{Response of Pion Wavefunctions to Hot Magnetized Nuclear Medium}
\begin{figure*}
	\centering

\begin{minipage}[c]{0.98\textwidth}
\hfill
\hspace{2pt}%
\begin{minipage}[c]{7.0cm}
    \centering
    $\rho_B=0$ \\[-2pt]
    
\end{minipage}
\hfill
\hspace{-2pt}%
\begin{minipage}[c]{7.0cm}
    \centering
    $\rho_B=\rho_0$ \\[-2pt]
    
\end{minipage}

    (a)\includegraphics[width=7.5cm,clip,trim=1.2cm 9.2cm 1.5cm 9cm]{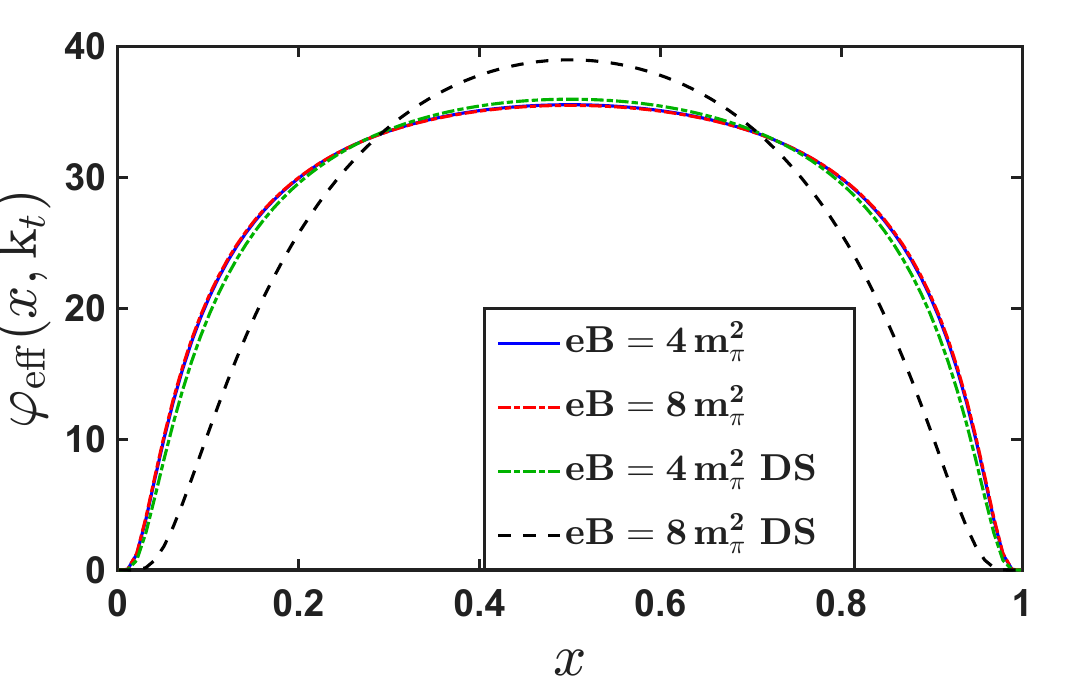}
    (b)\includegraphics[width=7.5cm,clip,trim=1.2cm 9.2cm 1.5cm 9cm]{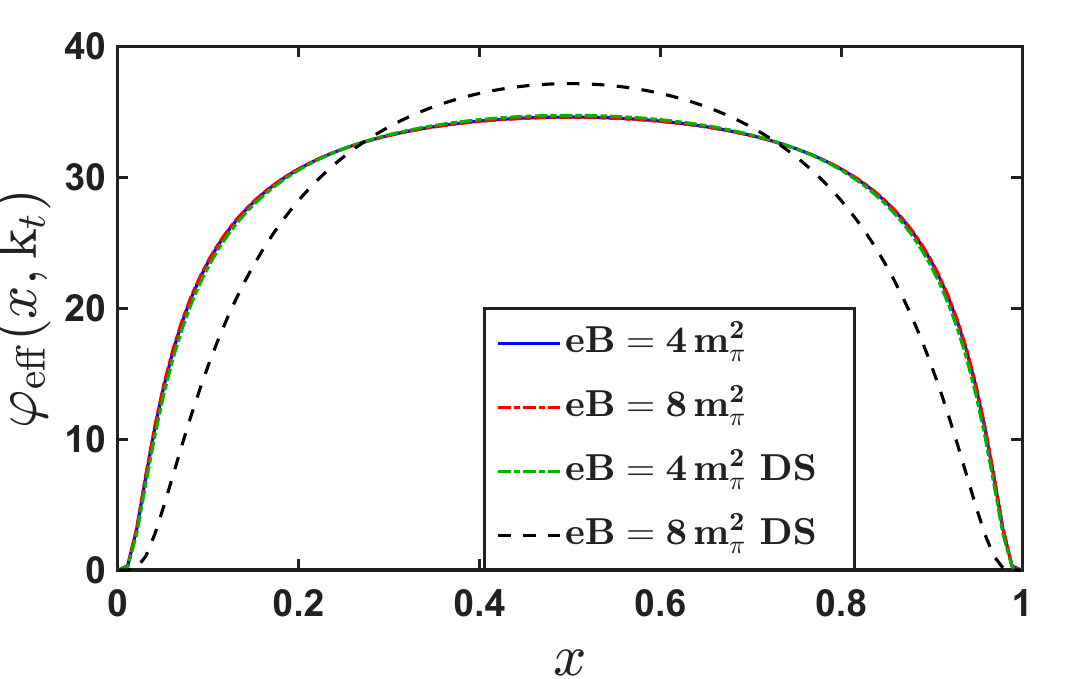}

\end{minipage}

\caption{LCWFs plotted as a function of the longitudinal momentum fraction $x$ at a fixed transverse momentum $\mathrm{k}_t = 0.2~\mathrm{GeV}$ for magnetic field strengths $eB = 4\,m_{\pi}^{2}$ and $eB = 8\,m_{\pi}^{2}$, with and without the inclusion of Dirac sea contributions, at temperature $T = 150~\mathrm{MeV}$ and isospin asymmetry $\eta = 0$. Subplots (a) and (b) correspond to baryon densities $\rho_B = 0$ and $\rho_0$, respectively.}
\label{fig:LCWF_DS}

\end{figure*}
\begin{figure*}
	\centering
\begin{minipage}[c]{0.98\textwidth}
\hfill
\hspace{2pt}%
\begin{minipage}[c]{7.0cm}
    \centering
     $\mathrm{\eta=0.3}$ \\[-2pt]
    
\end{minipage}
\hfill
\hspace{-2pt}%
\begin{minipage}[c]{7.0cm}
    \centering
    $\mathrm{\eta=0}$ \\[-2pt]
    
\end{minipage}
    (a)\includegraphics[width=7.5cm,clip,trim=1.2cm 9.2cm 1.5cm 9cm]{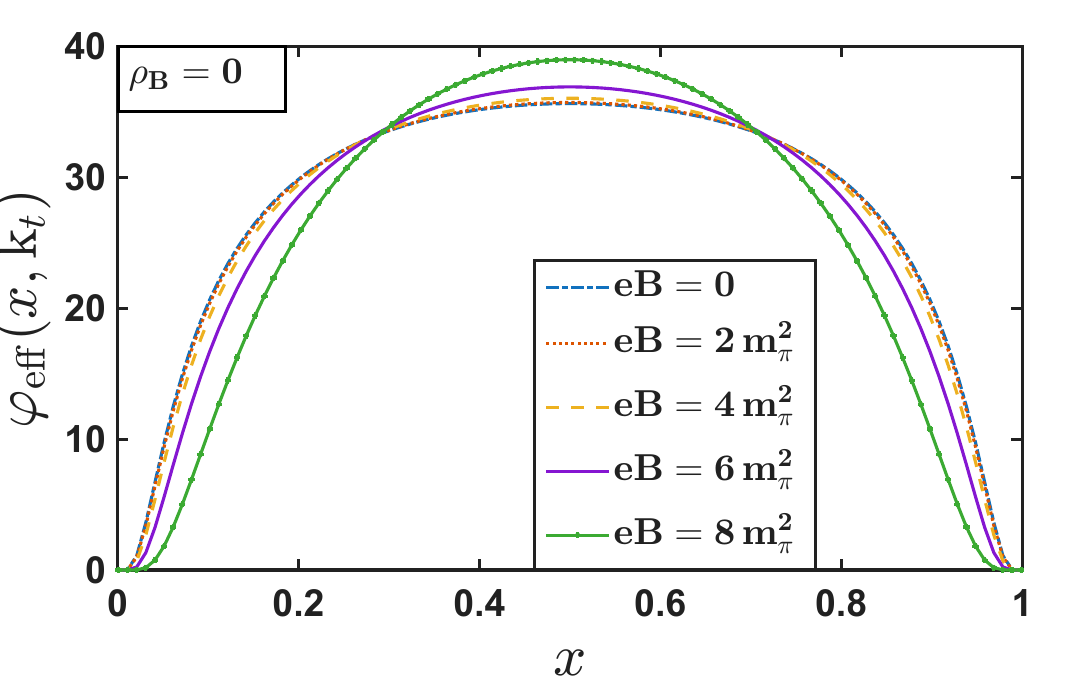}
    (b)\includegraphics[width=7.5cm,clip,trim=1.2cm 9.2cm 1.5cm 9cm]{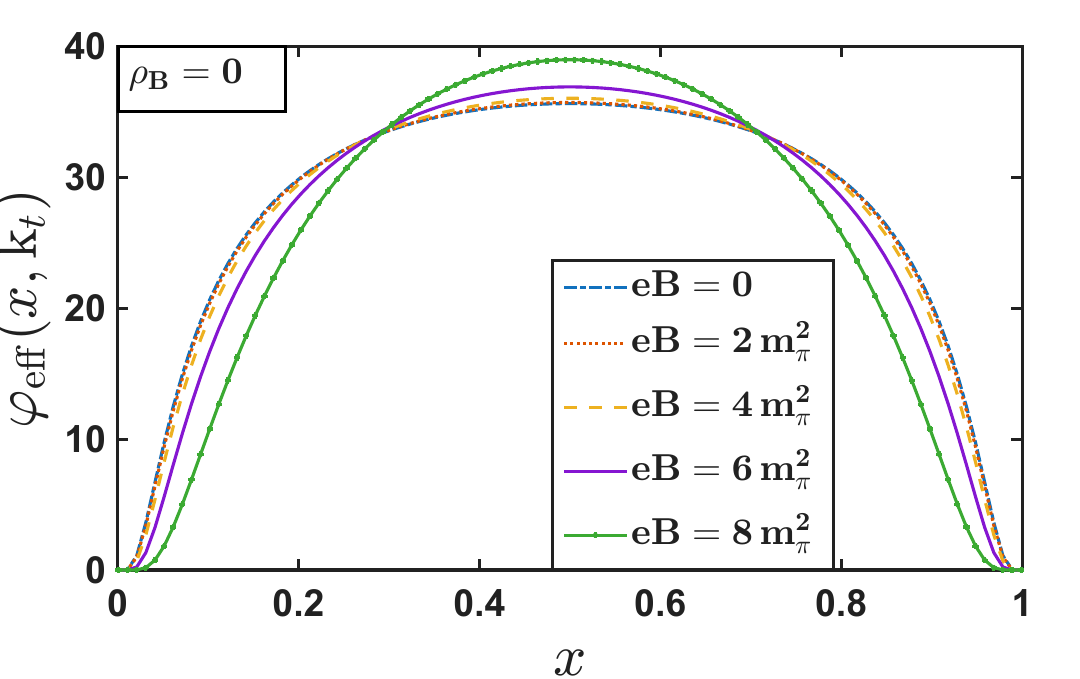}
    (c)\includegraphics[width=7.5cm,clip,trim=1.2cm 9.2cm 1.5cm 9cm]{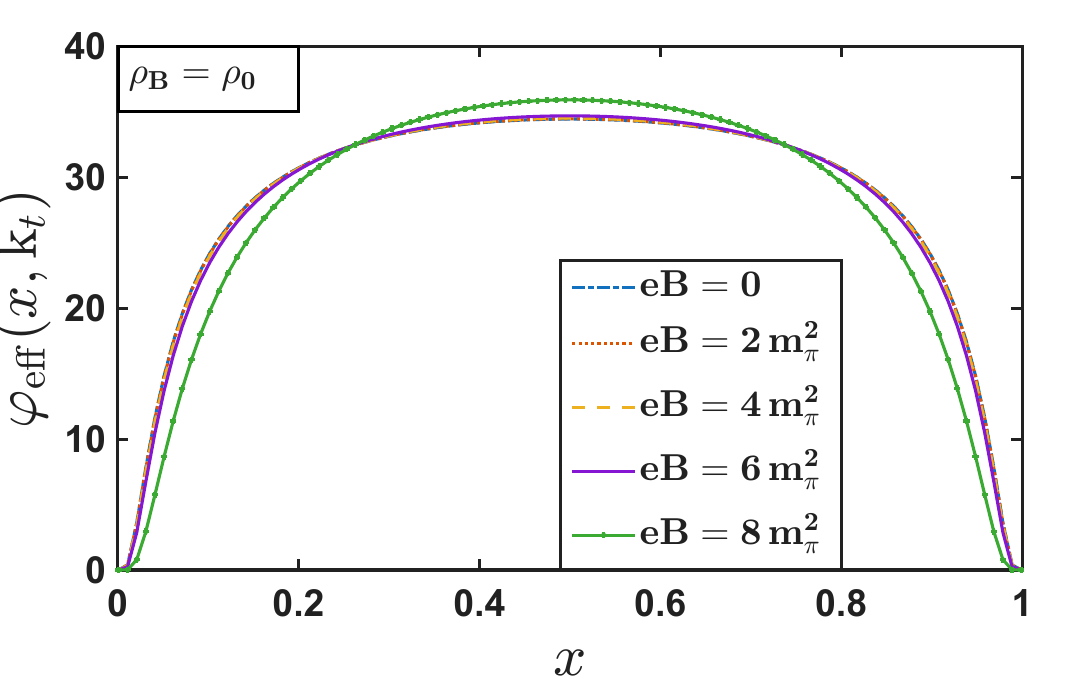}
    (d)\includegraphics[width=7.5cm,clip,trim=1.2cm 9.2cm 1.5cm 9cm]{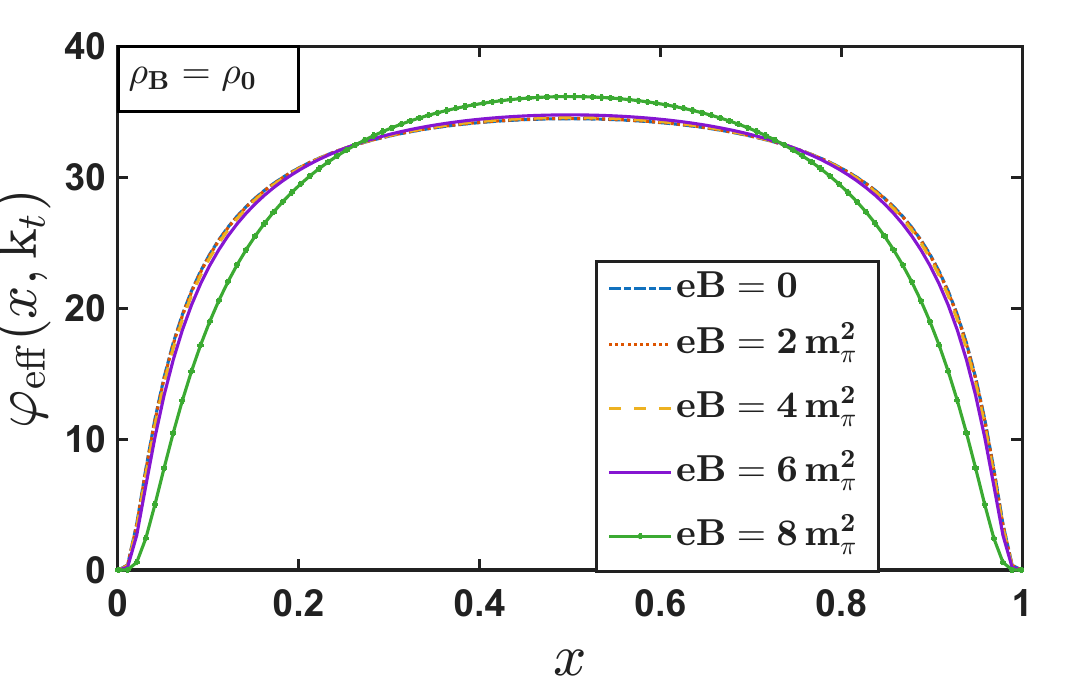}
    (e)\includegraphics[width=7.5cm,clip,trim=1.2cm 9.2cm 1.5cm 9cm]{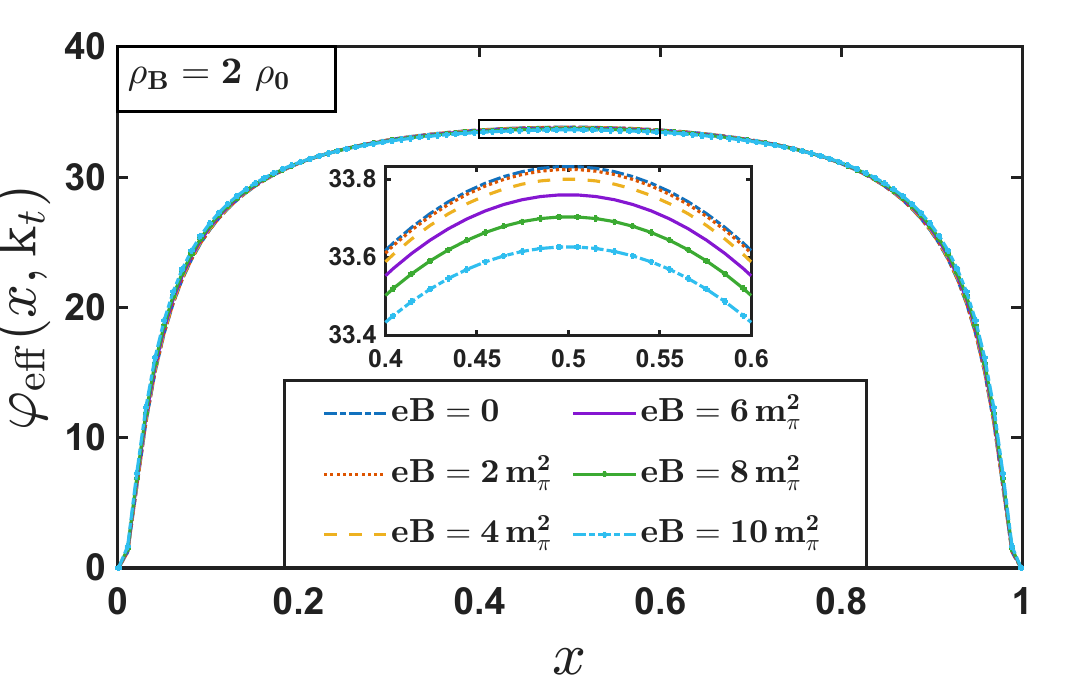}   (f)\includegraphics[width=7.5cm,clip,trim=1.2cm 9.2cm 1.5cm 9cm]{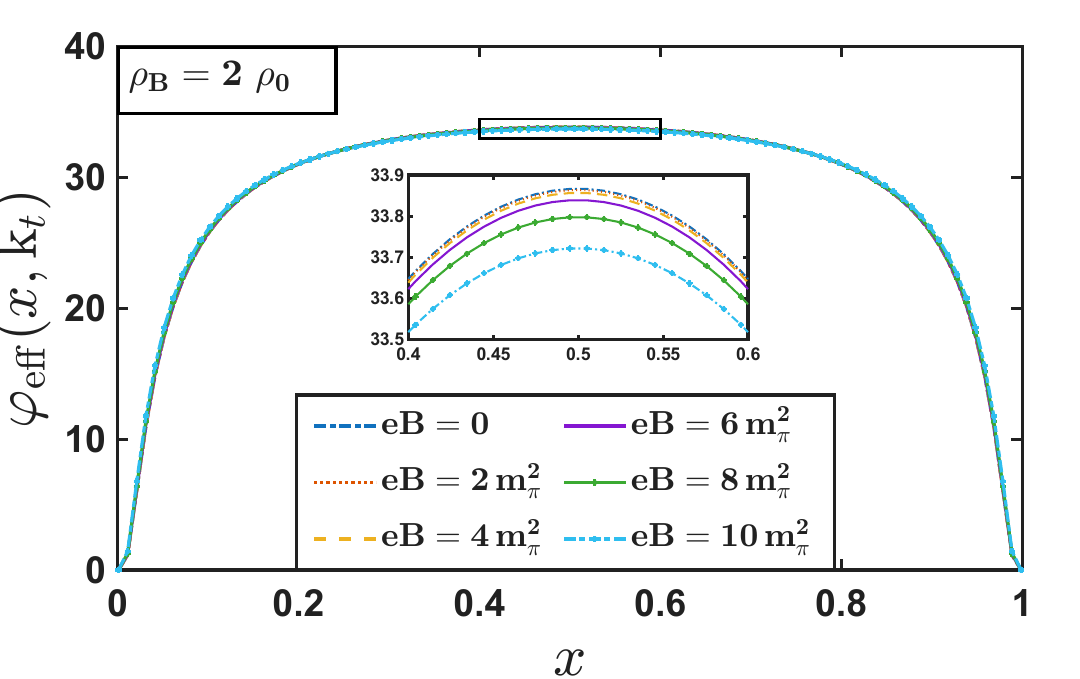}
\end{minipage}
\caption{LCWFs plotted as a function of the longitudinal momentum fraction $x$ at transverse momentum $\mathrm{k}_t=0.2\ \mathrm{GeV}$  for temperature $T = 50\,\mathrm{MeV}$, including Dirac sea contributions. The results are shown for various magnetic field strengths ranging from $eB = 0$ to $eB = 10\,m_{\pi}^2$. The left column (a, c, e) corresponds to symmetric nuclear matter ($\mathrm{\eta=0}$), while the right column (b, d, f) represents asymmetric nuclear matter ($\mathrm{\eta=0.3}$). The results are presented for different baryon densities: $\rho_B = 0$ (a, b), $\rho_0$ (c, d), and $2\rho_0$ (e, f).}
\label{fig:LCWF_T50}
\end{figure*}
\begin{figure*}
	\centering

\begin{minipage}[c]{0.98\textwidth}
\hfill
\hspace{2pt}%
\begin{minipage}[c]{7.0cm}
    \centering
     $\mathrm{\eta=0.3}$ \\[-2pt]
    
\end{minipage}
\hfill
\hspace{-2pt}%
\begin{minipage}[c]{7.0cm}
    \centering
    $\mathrm{\eta=0}$ \\[-2pt]
    
\end{minipage}

    (a)\includegraphics[width=7.5cm,clip,trim=1.2cm 9.2cm 1.5cm 9cm]{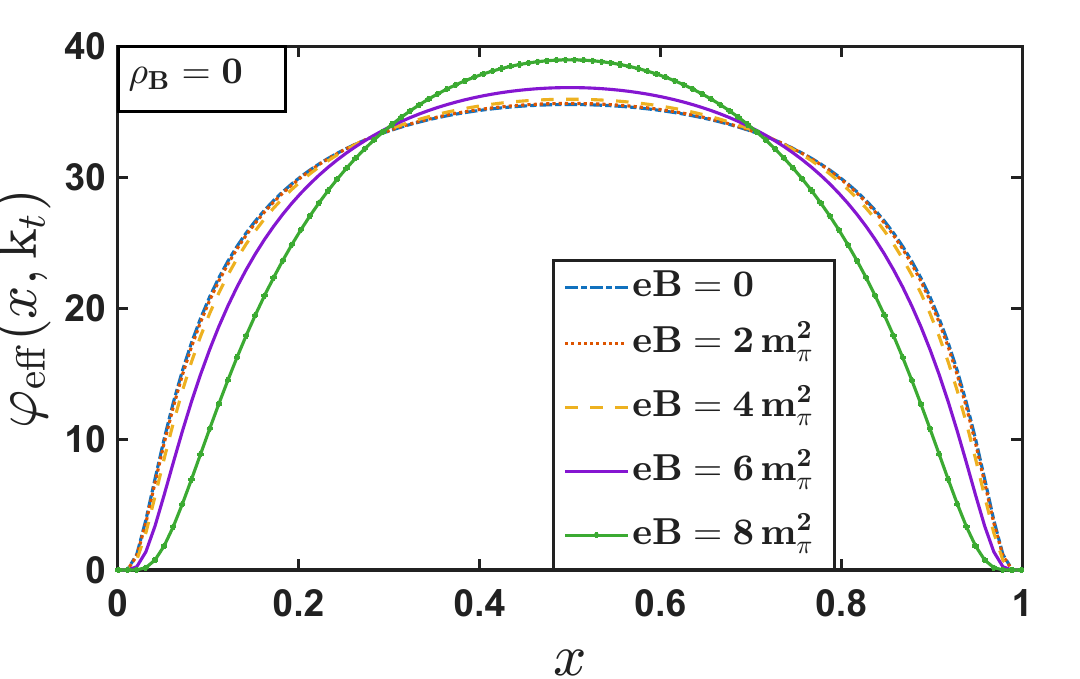}
    (b)\includegraphics[width=7.5cm,clip,trim=1.2cm 9.2cm 1.5cm 9cm]{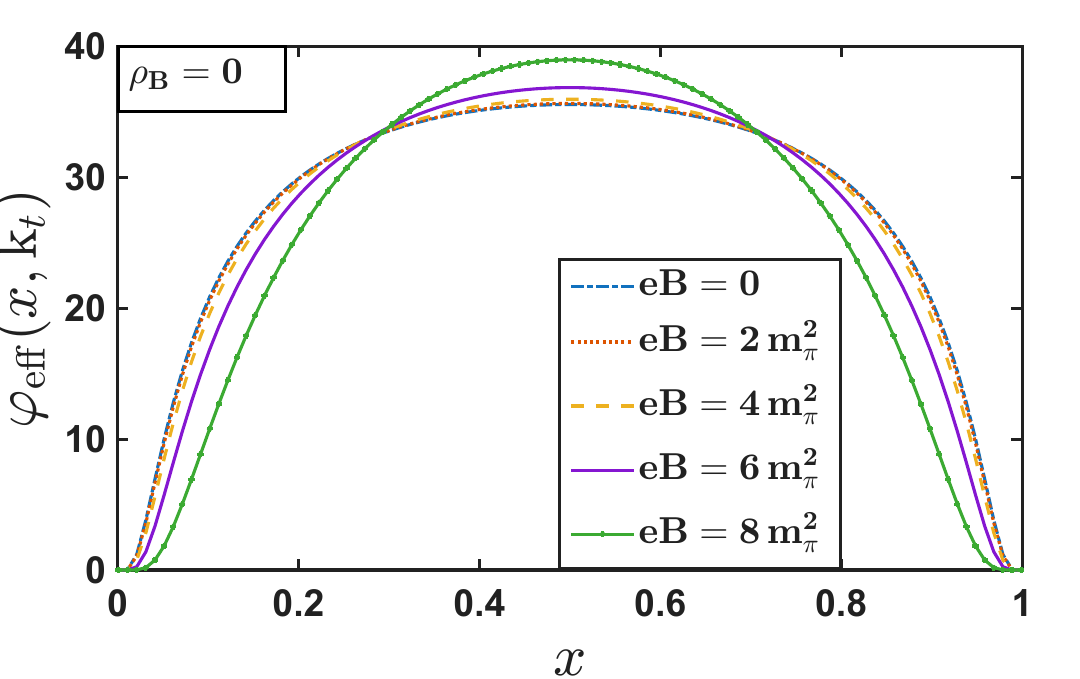}
    (c)\includegraphics[width=7.5cm,clip,trim=1.2cm 9.2cm 1.5cm 9cm]{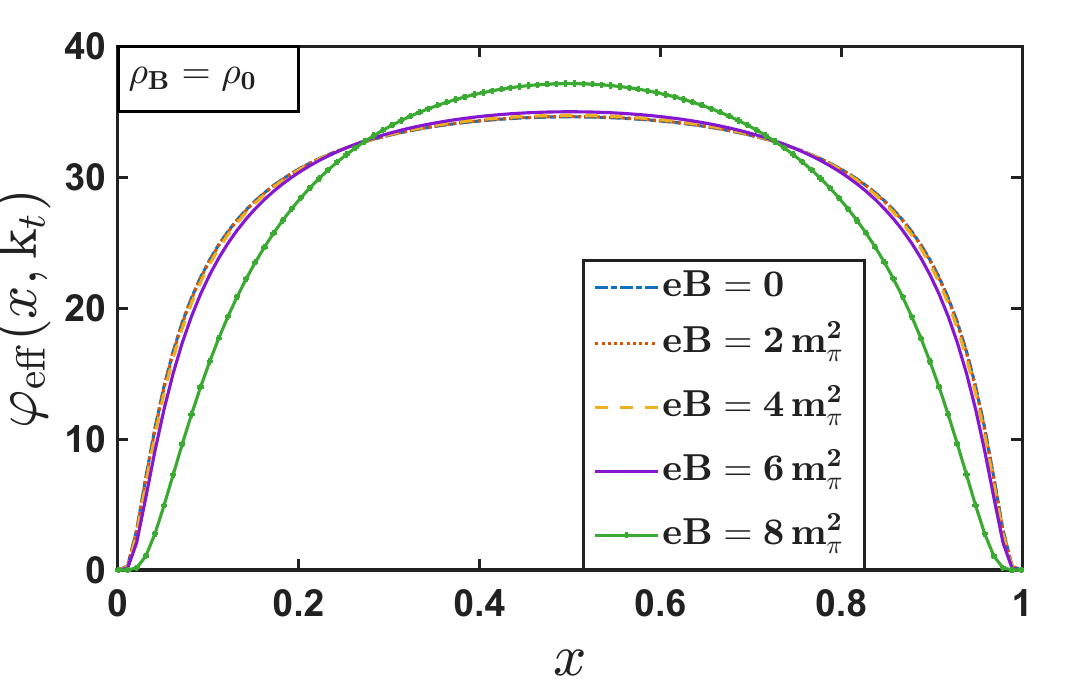}
    (d)\includegraphics[width=7.5cm,clip,trim=1.2cm 9.2cm 1.5cm 9cm]{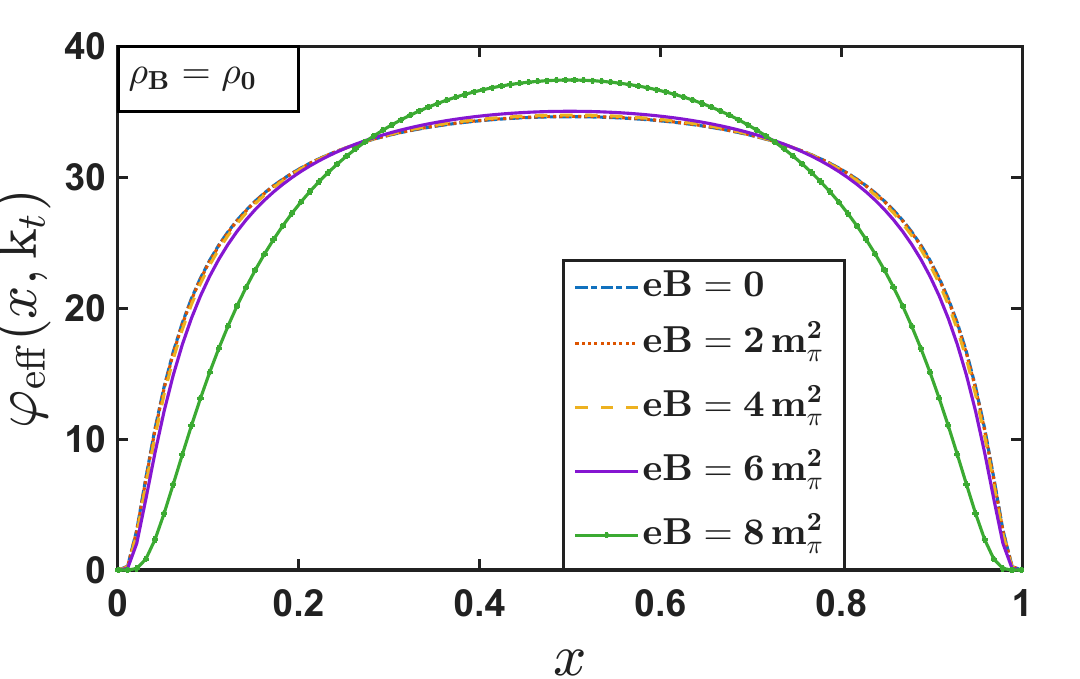}
    (e)\includegraphics[width=7.5cm,clip,trim=1.2cm 9.2cm 1.5cm 9cm]{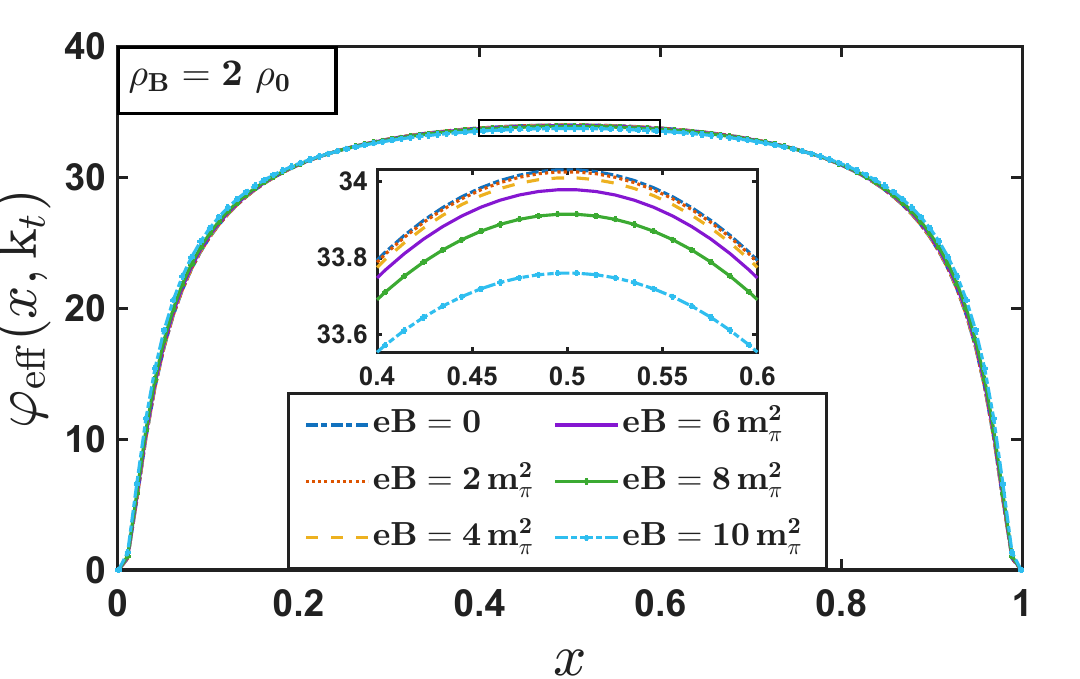}
    (f)\includegraphics[width=7.5cm,clip,trim=1.2cm 9.2cm 1.5cm 9cm]{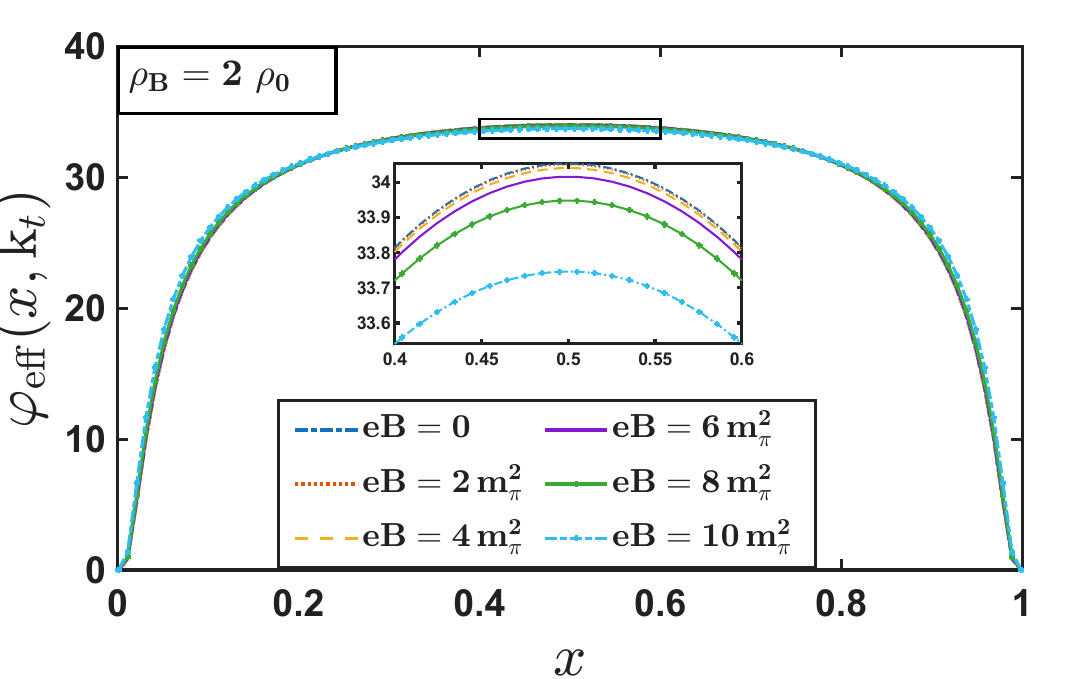}

\end{minipage}

\caption{Variation of the LCWFs with longitudinal momentum fraction $x$ at $\mathrm{k}_t=0.2\ \mathrm{GeV}$ for $T = 150\,\mathrm{MeV}$. All other conditions remain identical to those in Fig.~\ref{fig:LCWF_T50}.}
\label{fig:LCWF_T150}

\end{figure*}
\par The quark mass acts as the primary input in the LCQM, and hence any medium modification can be transmitted to LCWFs through $m_q^*$. Since the LCWFs depend explicitly on the quark mass, in-medium variations are clearly reflected in the pion wave functions. Figure~\ref{fig:LCWF_DS} illustrates the variation of the pion momentum space wave function $\varphi_{\text{eff}}(x,\mathrm{k}_t)$ as a function of the longitudinal momentum fraction $x$ at a fixed transverse momentum $\mathrm{k}_t = 0.2\ \mathrm{GeV}$, under different medium conditions. The subfigures (a) and (b) correspond to baryon densities $\mathrm{\rho_B = 0}$ and $\rho_0$, respectively, where $\varphi_{\text{eff}}(x,\mathrm{k}_t)$ is evaluated for both with and without the inclusion of DS contributions for magnetic field strengths $\mathrm{eB = 4 \ m_\pi^2}$ and $\mathrm{eB = 8\ m_\pi^2}$. It is observed that when DS contributions are ignored, the modification of LCWFs with magnetic field is negligible, consistent with the response of the scalar fields and the effective quark masses. However, when DS effects are included, the variation becomes significant, indicating the crucial role of vacuum contributions. Therefore, the subsequent analysis focuses merely on the case where DS effects are incorporated.
\par The variations in $\varphi_{\text{eff}}(x,\mathrm{k}_t)$ under different conditions of magnetic field, $\mathrm{\rho_B}$, and $\mathrm{\eta}$ at temperature $\mathrm{T = 50\ MeV}$ are shown in Fig.~\ref{fig:LCWF_T50}. At lower baryon densities, such as $\mathrm{\rho_B = 0}$ and $\mathrm{\rho_B = \rho_0}$, the peak of the distribution enhances around $x \sim 0.5$, accompanied by suppression at the end points. This behavior indicates a redistribution of momentum favoring more symmetric quark–antiquark configurations, which can be associated with an increase in the effective binding within the pion. However, this trend is reversed at higher density $\mathrm{\rho_B = 2\rho_0}$, where a suppression in the amplitude is observed. The reduction in the peak with increasing $\mathrm{\rho_B}$ can be associated with the decrease in $m_q^*$ and the partial restoration of chiral symmetry, which weakens the binding between quark and antiquark. Consequently, the wave function becomes more spread out, indicating a reduction in the probability of symmetric momentum sharing. The variations with respect to $\mathrm{\eta}$ and temperature are comparatively small at lower densities. It is also observed that these medium effects do not alter the symmetric nature of the pion wave function about $x = 0.5$. A similar behaviour is exhibited at $\mathrm{T = 150\ MeV}$, as shown in Fig.~\ref{fig:LCWF_T150}. Overall, the baryon density predominantly determines the response of the wave function, with magnetic field and temperature contributing secondary effects.
\subsection{Vacuum and in-medium Parton Distribution Functions}
\par The PDFs are evaluated from the LCWFs, which explicitly depend on the effective quark mass $m_q^*$. Consequently, modifications of $m_q^*$ induced by the surrounding medium are directly reflected in the resulting PDFs. In the vacuum, the pion PDF is symmetric about $x = 0.5$, indicating equal sharing of longitudinal momentum fraction between the quark and antiquark. It is also characterized by a peak around $x \sim 0.5$ and vanishing behavior at the boundaries ($x \to 0$ and $x \to 1$), consistent with the valence quark picture. These features provide a baseline for understanding how the internal momentum distribution of the pion is modified under different medium conditions.
\par
\begin{figure*}
	\centering

\begin{minipage}[c]{0.98\textwidth}
\hfill
\hspace{2pt}%
\begin{minipage}[c]{7.0cm}
    \centering
     $\mathrm{\eta=0}$ \\[-2pt]
    
\end{minipage}
\hfill
\hspace{-2pt}%
\begin{minipage}[c]{7.0cm}
    \centering
    $\mathrm{\eta=0.3}$ \\[-2pt]
    
\end{minipage}

    (a)\includegraphics[width=7.5cm,clip,trim=1.2cm 9.2cm 1.5cm 9cm]{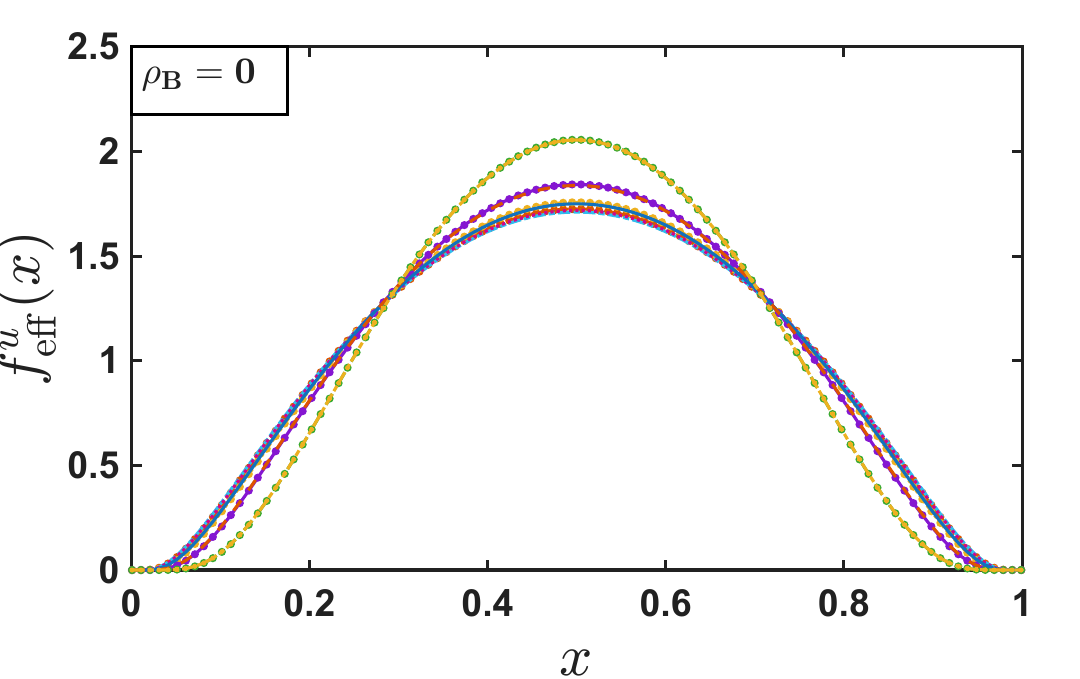}
    (b)\includegraphics[width=7.5cm,clip,trim=1.2cm 9.2cm 1.5cm 9cm]{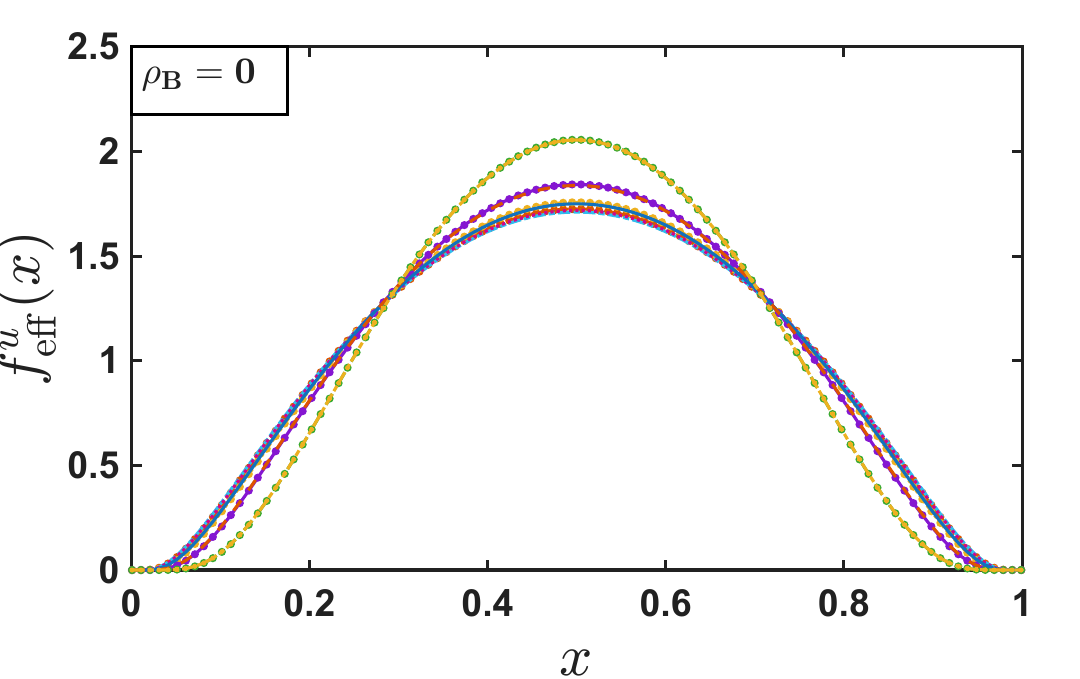}
    (c)\includegraphics[width=7.5cm,clip,trim=1.2cm 9.2cm 1.5cm 9cm]{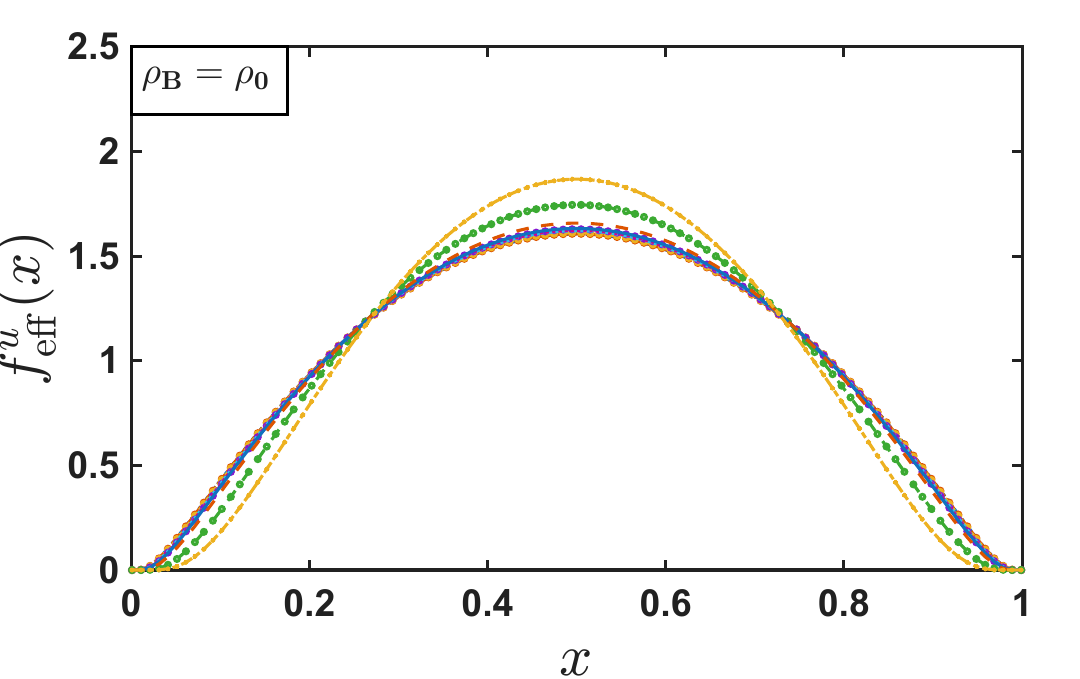}
    (d)\includegraphics[width=7.5cm,clip,trim=1.2cm 9.2cm 1.5cm 9cm]{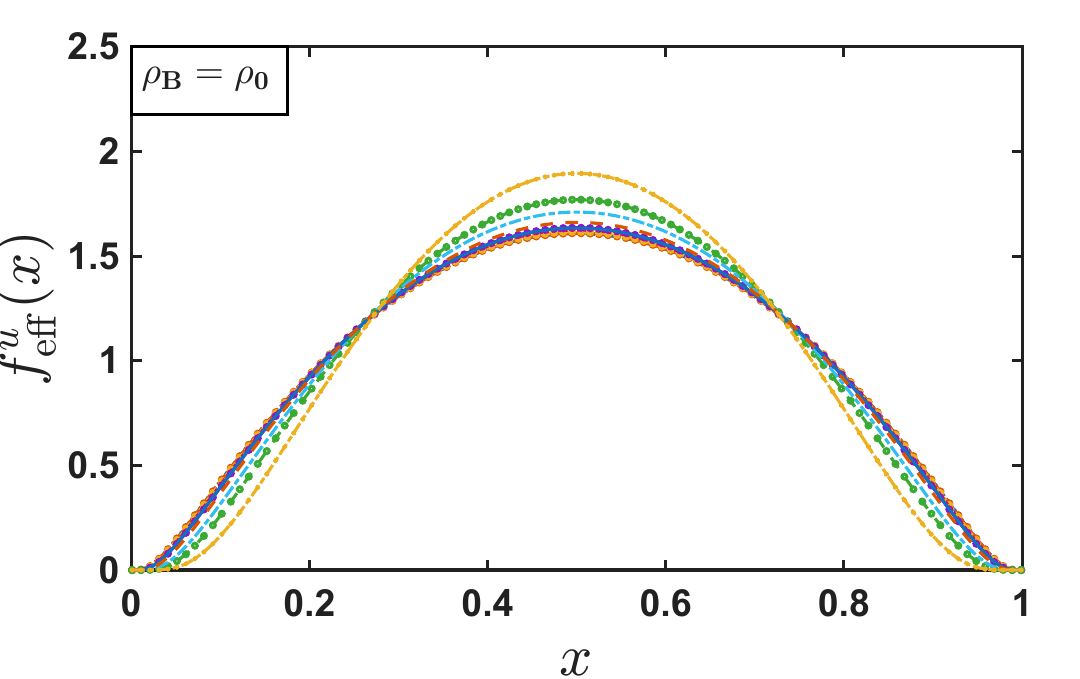}
\end{minipage}
\includegraphics[width=0.7\linewidth,clip,trim=3.5cm 11.8cm 2.4cm 14.3cm]{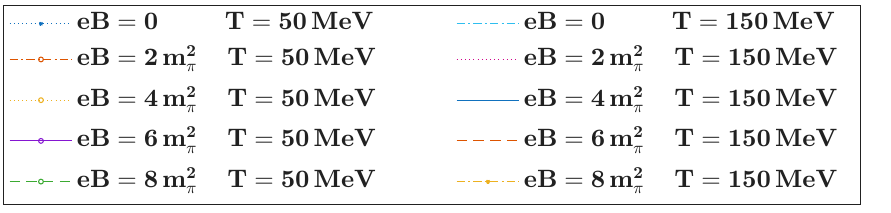}       
    \caption{The unpolarized PDF for the $u$ quark is plotted as a function of the longitudinal momentum fraction $x$. Subplots (a) and (b) correspond to baryon density $\mathrm{\rho_B = 0}$, while (c) and (d) represent $\mathrm{\rho_B = \rho_0}$. The results are presented for various magnetic field strengths and temperatures $\mathrm{T = 50\ MeV}$ and $\mathrm{T = 150\ MeV}$, at fixed isospin asymmetry parameters $\mathrm{\eta = 0}$ (a, c) and $\mathrm{\eta = 0.3}$ (b, d).}
    \label{fig:pdf_temp_01}
\end{figure*}

\begin{figure*}
	\centering
\begin{minipage}[c]{0.98\textwidth}
\hfill
\hspace{2pt}%
\begin{minipage}[c]{7.0cm}
    \centering
     $\mathrm{\eta=0}$ \\[-2pt]
\end{minipage}
\hfill
\hspace{-2pt}%
\begin{minipage}[c]{7.0cm}
    \centering
    $\mathrm{\eta=0.3}$ \\[-2pt]
\end{minipage}
    (a)\includegraphics[width=7.5cm,clip,trim=1.2cm 9.2cm 1.5cm 9cm]{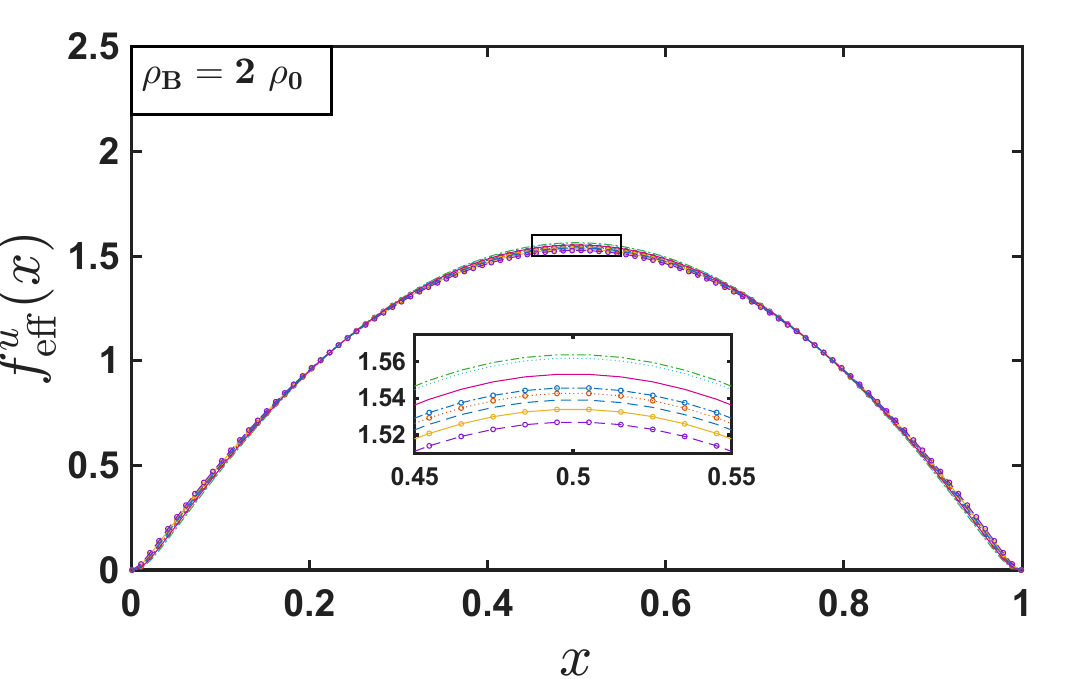}   (b)\includegraphics[width=7.5cm,clip,trim=1.2cm 9.2cm 1.5cm 9cm]{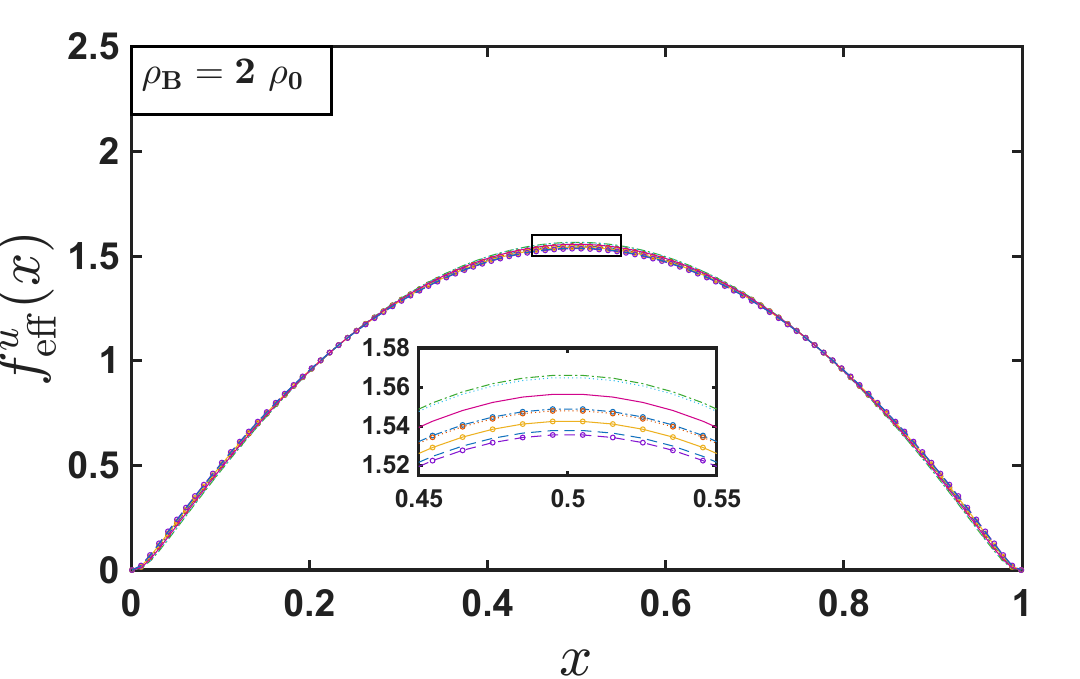}
\end{minipage}
\includegraphics[width=0.75\linewidth,clip,trim=3.5cm 
            13cm 5cm 14.4cm]{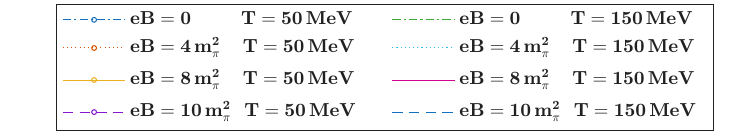}
    \caption{The unpolarized PDF for the $u$ quark is displayed with respect to  longitudinal momentum fraction $x$ for baryon density $\mathrm{\rho_B=2 \ \rho_0}$ with conditions similar to that of Fig.~\ref{fig:pdf_temp_01}}
    \label{fig:pdf_temp_2}
\end{figure*}

 In Fig.~\ref{fig:pdf_temp_01}, the variation of the effective unpolarized $u$-quark PDF as a function of longitudinal momentum fraction $x$ is illustrated for different magnetic field strengths $eB$ and temperature $\mathrm{T (MeV)}$ at fixed isospin asymmetry $\mathrm{\eta=0}$  (left panel) and $\mathrm{\eta=0.3}$ (right panel). The results are shown for $\mathrm{\rho_B = 0}$ (a, b) and $\mathrm{\rho_B = \rho_0}$ (c, d). It is observed that the amplitude around $x \sim 0.5$ increases with magnetic field in both cases, which can be attributed to the enhancement of effective quark mass due to magnetic catalysis, leading to stronger binding and a more localized momentum distribution. For $\mathrm{\rho_B = 0}$, no noticeable difference is observed between symmetric and asymmetric matter, as the curves overlap for the same magnetic field at both temperatures, indicating negligible isospin effects in the absence of medium interactions. However, at $\mathrm{\rho_B = \rho_0}$, a small difference emerges between the two cases due to the contribution of the $\mathrm{\delta}$ field in asymmetric matter, which slightly modifies $m_q^*$ of $u$ and $d$ quarks. This difference becomes marginally more visible at higher temperature (as shown in Fig.\ref{fig:pdf_temp_01}(d)), suggesting that thermal effects enhance the sensitivity of the system to isospin asymmetry. The corresponding results for $\mathrm{\rho_B = 2\rho_0}$ are shown in Fig.~\ref{fig:pdf_temp_2}, where the magnetic field dependence is reversed, with the decrease in amplitude as the magnetic field strength increases. This trend arises from the dominance of density effects, where the reduction in scalar fields leads to a decrease in $m_q^*$, resulting in scaling down and broadening of momentum distribution. Although the effect of $\mathrm{\eta}$ and $\mathrm{T (MeV)}$ is still present, but it appears less pronounced compared to $\mathrm{\rho_B = \rho_0}$. %because the overall changes induced by the magnetic field are comparatively weak.
%, not due to a reduction in isospin influence, but
%
%
%
%
%
%
%
%\par The PDFs are further shown in Figs.~\ref{fig:pdf_temp_01} and \ref{fig:pdf_temp_2} to investigate the effect of temperature, with $\mathrm{\eta}$ kept constant for each plot. The magnetic field dependence remains similar to the previous case. At $\mathrm{\rho_B = 0}$, the PDFs corresponding to different temperatures are observed to coincide, showing no significant variation. This is because, in the absence of medium effects, thermal contributions are not sufficient to modify the scalar fields and hence the effective quark mass remains nearly unchanged. As in the previous case, some variations appear at $\mathrm{\rho_B = \rho_0}$, but here the relative shift is more noticeable compared to the $\mathrm{\eta}$ dependence. This occurs because the temperature affects the scalar fields more effectively in the presence of finite density, leading to small but visible modifications in the quark mass and hence the PDF. The change becomes slightly more pronounced for $\mathrm{\eta = 0.3}$, as the combined effect of $\mathrm{\eta}$ and T, enhances the sensitivity of the system. The case of $\mathrm{\rho_B = 2\rho_0}$ follows a similar trend as discussed earlier, where the PDF shows an overall decrease in amplitude. Overall, the temperature effect on PDFs is relatively weak and becomes noticeable only for finite values of baryon density and isospin asymmetry
%
\subsection{Medium Effects on Pion Electromagnetic Form Factors}
%
%
%
%%
%
%
% \begin{figure*}
% 	\centering
% \begin{minipage}[c]{0.98\textwidth}
% \hfill
% \hspace{2pt}%
% \begin{minipage}[c]{7.0cm}
%     \centering
%      $\mathrm{T=50\ MeV}$ \\[-2pt]
    
% \end{minipage}
% \hfill
% \hspace{-2pt}%
% \begin{minipage}[c]{7.0cm}
%     \centering
%     $\mathrm{T=150\ MeV}$ \\[-2pt]
    
% \end{minipage}
%     (a)\includegraphics[width=7.5cm,clip,trim=1cm 9.2cm 1.5cm 9cm]{images/ff_eta_comb/rho_0/T_50.pdf}
%     (b)\includegraphics[width=7.5cm,clip,trim=1cm 9.2cm 1.5cm 9cm]{images/ff_eta_comb/rho_0/T_150.pdf}
%     (c)\includegraphics[width=7.5cm,clip,trim=1cm 9.2cm 1.5cm 9cm]{images/ff_eta_comb/rho_1/T_50.pdf}
%     (d)\includegraphics[width=7.5cm,clip,trim=1cm 9.2cm 1.5cm 9cm]{images/ff_eta_comb/rho_1/T_150.pdf}
%     (e)\includegraphics[width=7.5cm,clip,trim=1cm 9.2cm 1.5cm 9cm]{images/ff_eta_comb/rho_2/T_50.pdf}   (f)\includegraphics[width=7.5cm,clip,trim=1cm 9.2cm 1.5cm 9cm]{images/ff_eta_comb/rho_2/T_150.pdf}
% \end{minipage}
% \caption{EMFFs for the $u$ quark are illustrated with respect to the momentum transfer $\mathrm{Q^2}$. for baryon densities $\mathrm{\rho_B = 0}$ (a, b), $\mathrm{\rho_B = \rho_0}$ (c, d), and $\mathrm{\rho_B = 2\rho_0}$ (e, f). The outcomes are presented for different strengths of the magnetic field and isospin asymmetry parameters $\mathrm{\eta = 0}$ and $\mathrm{\eta = 0.3}$, at fixed temperatures $\mathrm{T = 50\ MeV}$ (a, c, e) and $\mathrm{T = 150\ MeV}$ (b, d, f).}
%     \label{fig:EMFF_eta}
% \end{figure*}
% %
%
%
%%
%
%
\begin{figure*}
	\centering
\begin{minipage}[c]{0.98\textwidth}
\hfill
\hspace{2pt}%
\begin{minipage}[c]{7.0cm}
    \centering
     $\mathrm{\eta=0}$ \\[-2pt]
    
\end{minipage}
\hfill
\hspace{-2pt}%
\begin{minipage}[c]{7.0cm}
    \centering
    $\mathrm{\eta=0.3}$ \\[-2pt]
    
\end{minipage}
    (a)\includegraphics[width=7.5cm,clip,trim=1cm 9.2cm 1.5cm 9cm]{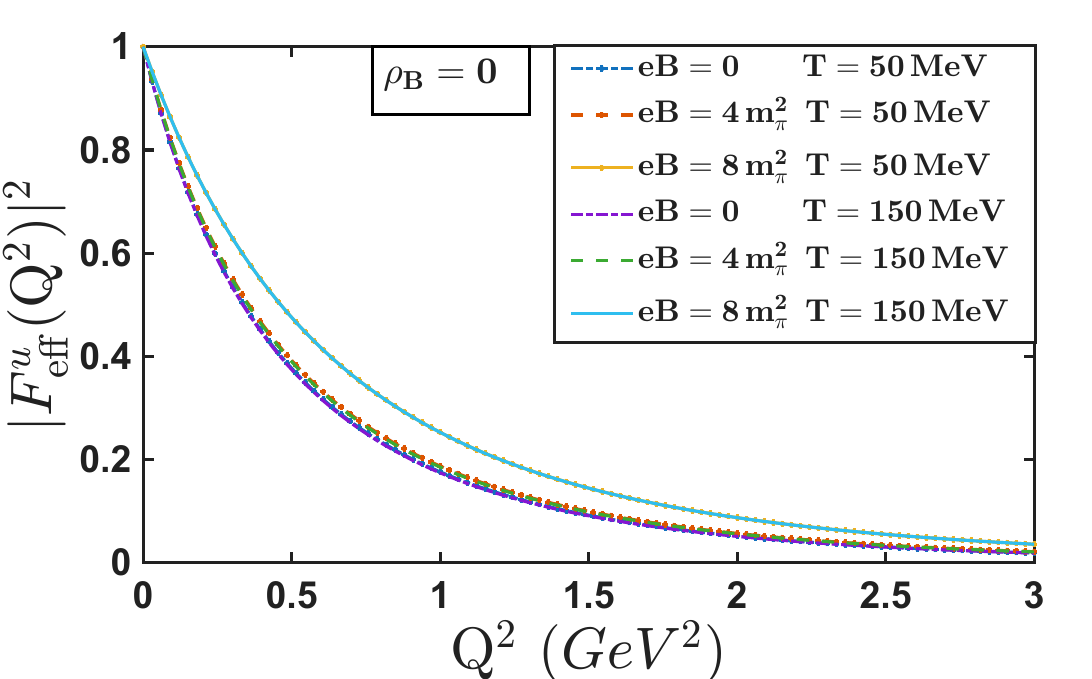}
    (b)\includegraphics[width=7.5cm,clip,trim=1cm 9.2cm 1.5cm 9cm]{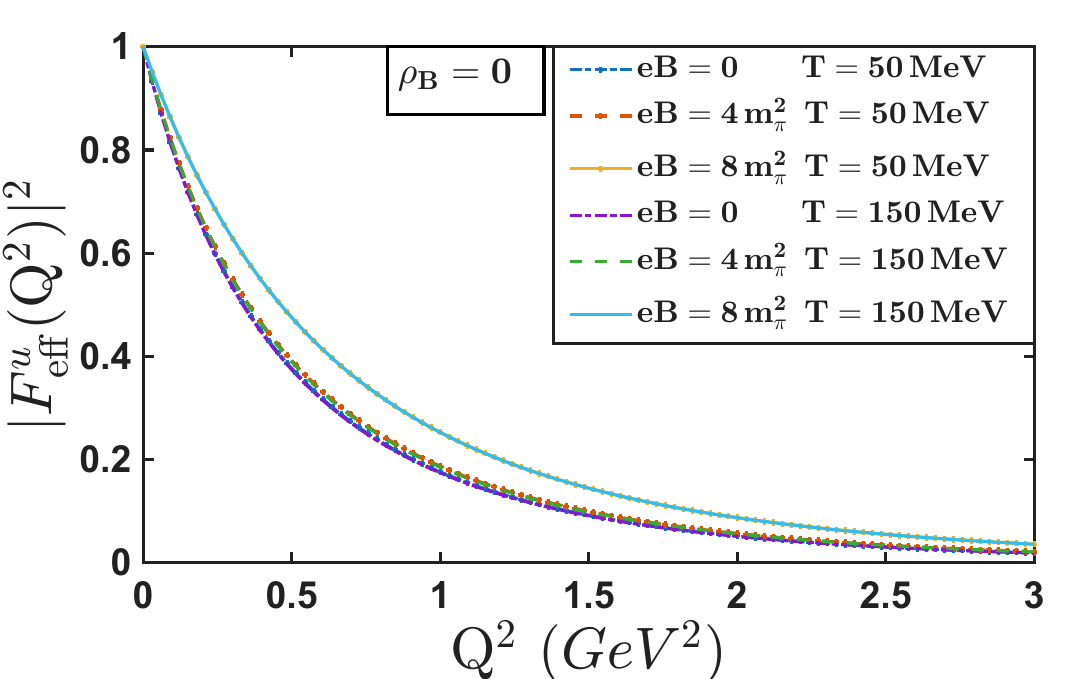}
    (c)\includegraphics[width=7.5cm,clip,trim=1cm 9.2cm 1.5cm 9cm]{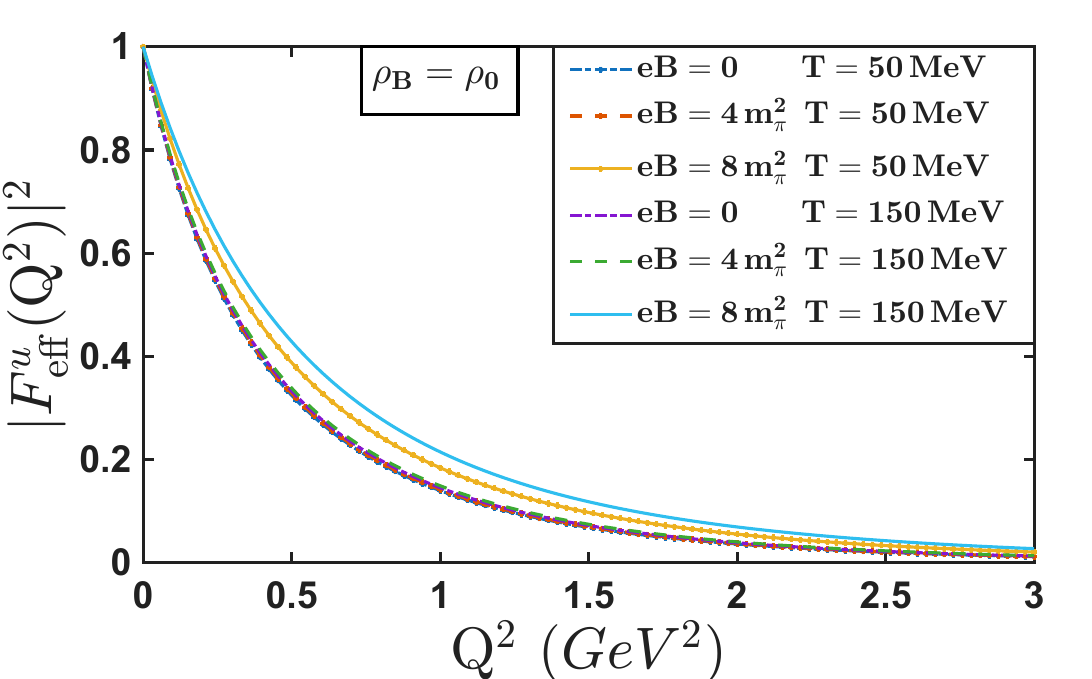}
    (d)\includegraphics[width=7.5cm,clip,trim=1cm 9.2cm 1.5cm 9cm]{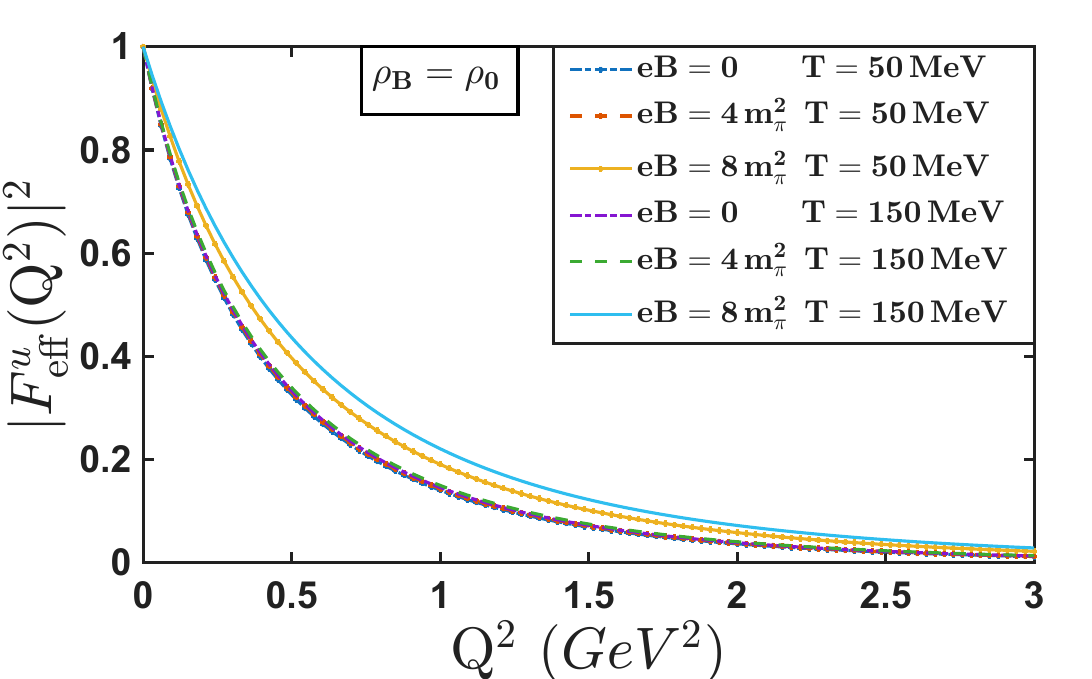}
    (e)\includegraphics[width=7.5cm,clip,trim=1cm 9.2cm 1.5cm 9cm]{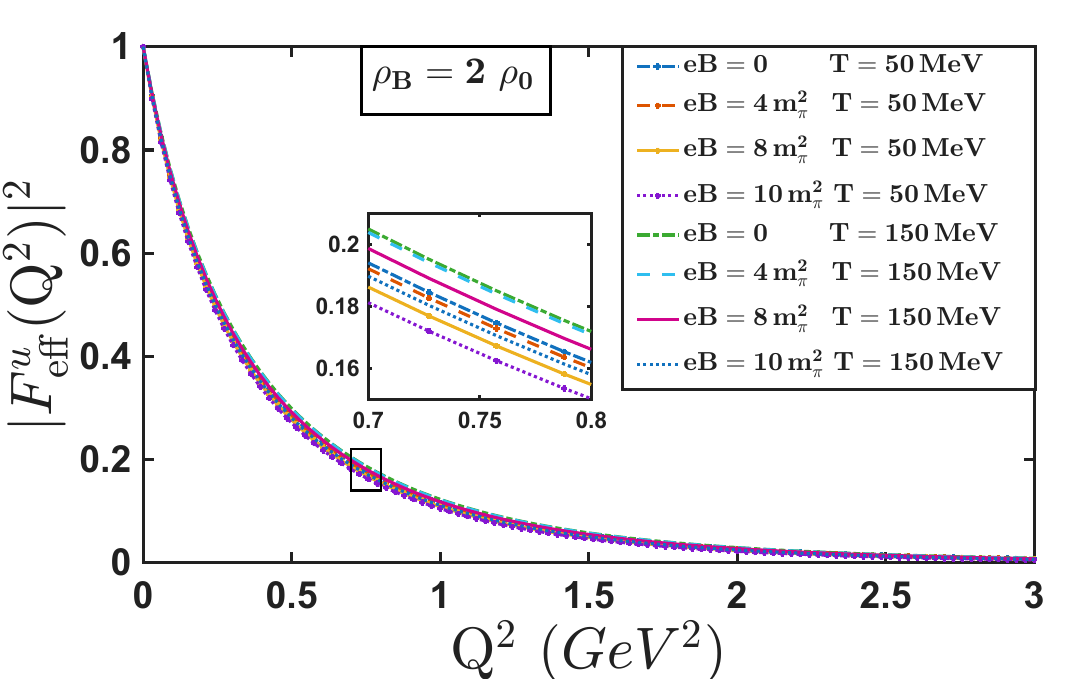}   (f)\includegraphics[width=7.5cm,clip,trim=1cm 9.2cm 1.5cm 9cm]{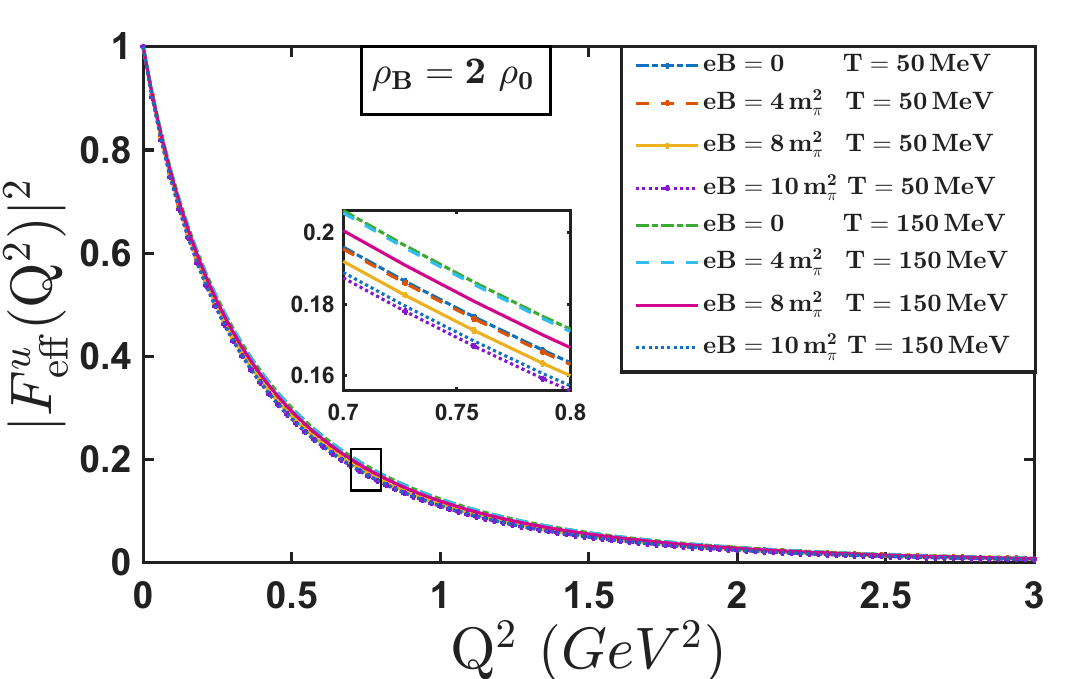}
\end{minipage}
 \caption{EMFFs corresponding to the $u$ quark are illustrated with respect to the momentum transfer $\mathrm{Q^2}$ at $\mathrm{\rho_B = 0}$ (a, b), $\mathrm{\rho_B = \rho_0}$ (c, d), and $\mathrm{\rho_B = 2\rho_0}$ (e, f). The calculations are performed for a range of magnetic field strengths and temperatures $\mathrm{T = 50\ MeV}$ and $\mathrm{T = 150\ MeV}$, while maintaining fixed values of $\mathrm{\eta = 0}$ (a, c, e) and $\mathrm{\eta = 0.3}$ (b, d, f).}
    \label{fig:EMFF_temp}
\end{figure*}
\par 
%EMFFs are evaluated from the LCWFs, and therefore the influence of the medium is included through $m_q^*$. Since the LCWFs depend explicitly on the quark mass, any modification in the medium is directly reflected in the EMFFs. 
The impact of the medium on the hadron structure is further examined through the EMFFs.
In the vacuum, the pion EMFF describes the spatial distribution of charge and decreases smoothly with increasing momentum transfer $\mathrm{Q^2 (GeV^2)}$, providing a baseline to study medium-induced modifications.
%$\mathrm{T (MeV)}$

%
\par In Fig.~\ref{fig:EMFF_temp}, the effective EMFFs of the $u$-quark are presented as a function of the momentum transfer $\mathrm{Q^2 (GeV^2)}$ for three different magnetic field strengths $\mathrm{eB=\{0, 4m_\pi^2, 8m_\pi^2\}}$ and two discrete values of temperatures $\mathrm{T = 50}$ and $\mathrm{150\ MeV}$ at fixed isospin asymmetry $\mathrm{\eta=0}$ (left panel) and $\mathrm{0.3}$ (right panel). The results are shown for baryon densities $\mathrm{\rho_B = 0}$ (a, b), $\mathrm{\rho_B = \rho_0}$ (c, d), and $\mathrm{\rho_B = 2\rho_0}$ (e, f). At lower baryon densities, $\mathrm{\rho_B = 0}$ and $\mathrm{\rho_B = \rho_0}$, the EMFFs show a rise with increasing magnetic field strength. This enhancement can be associated with the increase in effective quark mass under magnetic field effects, which leads to a more compact charge distribution and hence larger form factor values. For $\mathrm{\rho_B = 0}$,  temperature dependence is negligible for both symmetric and asymmetric nuclear matter. However, at $\mathrm{\rho_B = \rho_0}$, a slight separation between the distributions becomes visible between $\mathrm{T = 50}$ and $\mathrm{150\ MeV}$ , which becomes marginally more pronounced at $\mathrm{eB= 8m_\pi^2}$. The effect of medium asymmetry is also seen at saturation density of the nuclear medium (as shown in Fig.~\ref{fig:EMFF_temp}(d)), which leads to greater separation between the distributions of $\mathrm{eB= 8m_\pi^2}$ and $\mathrm{eB= 4m_\pi^2}$ than the case of symmetric matter.
In contrast, at higher density $\mathrm{\rho_B = 2\rho_0}$, the magnetic field dependence shows an opposite trend, with the form factors decreasing as the magnetic field increases. These trends are consistent with those observed in the PDFs discussed earlier, indicating a similar underlying response of the pion structure to medium modifications.

%\par Figure~\ref{fig:EMFF_temp} presents the EMFF for different temperatures at fixed $\mathrm{\eta}$, in order to examine the temperature dependence along with magnetic field effects. The overall behavior with respect to magnetic field and baryon density remains consistent with that observed in Fig.~\ref{fig:EMFF_eta}. At $\mathrm{\rho_B = 0}$, the curves corresponding to different temperatures coincide for a given magnetic field, indicating negligible thermal effects in the absence of medium interactions. However, at finite densities, the curves corresponding to different temperatures show a noticeable shift, reflecting the increasing influence of temperature in the presence of a medium, as discussed earlier. This behavior is also consistent with the trends observed in the PDFs (see Figs. \ref{fig:pdf_temp_01} and \ref{fig:pdf_temp_2}).
%
%
%
%
\section{Summary and Conclusions}
\label{sec:Summary and Conclusions}
\par In this work, we have investigated the modification of pion structure in magnetized nuclear matter at finite temperature within a unified framework combining the chiral SU(3) quark mean field model and the light-cone quark model. The in-medium quark masses obtained from the scalar fields were used as inputs to compute the light-cone wave functions (LCWFs), from which the parton distribution functions (PDFs) and electromagnetic form factors (EMFFs) of the pion were evaluated under different thermodynamic conditions.
\par Our analysis shows that the inclusion of Dirac Sea (DS) contributions plays a crucial role in determining the response of the system to the magnetic field. While the scalar fields exhibit minimal sensitivity to the magnetic field in the absence of DS effects, their behavior is significantly modified upon inclusion of DS contributions, leading to opposite trends at low and high baryon densities. This behavior is associated with inverse magnetic catalysis in dense matter. The corresponding modifications in scalar fields are reflected in the in-medium quark masses, which in turn influence the pion structure.
\par The LCWFs, PDFs, and EMFFs exhibit consistent trends across the considered parameters. At lower baryon densities, these observables increase with magnetic field strength, whereas at higher density ($\rho_B = 2\rho_0$), a decreasing behavior is observed. This can be attributed to the interplay between magnetic field effects and density-driven modifications of the effective quark mass. At zero baryon density, medium effects are negligible, while at finite densities, temperature effects are found to be more prominent than isospin asymmetry in influencing the pion structure.
\par Overall, the results demonstrate that the properties of the pion can deviate significantly from their vacuum values in a magnetized medium. Such studies are relevant for understanding hadronic matter under extreme conditions, as encountered in heavy-ion collisions and compact astrophysical objects. The present framework can be extended to investigate other in-medium properties of hadrons, providing further insights into the behavior of strongly interacting matter.
\section*{Acknowledgement}
{H.D. would like to thank the Science and Engineering Research Board, Anusandhan
National Research Foundation (ANRF), Government of India under the SERB-POWER Fellowship scheme (Ref No. SPF/2023/000116) for financial support. A.K. sincerely acknowledges Anusandhan-National Research Foundation (ANRF), Government of India for
funding of the research project under the Science and Engineering Research Board-Core
Research Grant (SERB-CRG) scheme (File No. CRG/2023/000557).} 
%-----------------------------------------------------------
%
%
%
%
%
%Reference

%\bibliographystyle{apsrev4}
\bibliography{references}

\end{document}